\documentclass[preprints,article,accept,pdftex,moreauthors]{Definitions/mdpi} 

\firstpage{1} 
\pubvolume{1}
\issuenum{1}
\articlenumber{0}
\pubyear{2022}
\copyrightyear{2022}
\externaleditor{} 
\datereceived{} 
\dateaccepted{} 
\datepublished{} 
\hreflink{https://doi.org/} 
\pdfoutput=1

\usepackage{siunitx}

\newcommand*\revSL[1]{\textcolor{black}{#1}}
\renewcommand*\hl[1]{\textcolor{black}{#1}}

\Title{Biophysical Modeling of SARS-CoV-2 Assembly: Genome Condensation and Budding}

\Author{\hl{Siyu Li} 
 $^{1}$ and Roya Zandi $^{2}$}

\address{%
$^{1}$ \quad Songshan Lake Materials Laboratory, {Dongguan} 
 523808, China\\
$^{2}$ \quad Department of Physics and Astronomy, University of California Riverside, Riverside, CA 92521, USA}

\abstract{The COVID-19 pandemic caused by the Severe Acute Respiratory Syndrome Coronavirus 2 (SARS-CoV-2) has spurred unprecedented and concerted worldwide research to curtail and eradicate this pathogen. SARS-CoV-2 has four structural proteins: Envelope (E), Membrane (M), Nucleocapsid (N), and Spike (S), which self-assemble along with its RNA into the infectious virus by budding from intracellular lipid membranes.  In this paper, we develop a model to explore the mechanisms of RNA condensation by structural proteins, protein oligomerization and cellular membrane--protein interactions that control the budding process and the ultimate virus structure. Using molecular dynamics simulations, we have deciphered how the positively charged N proteins interact and condense the very long genomic RNA resulting in its packaging by a lipid envelope decorated with structural proteins inside a host cell.  Furthermore, considering the length of RNA and the size of the virus, we find that the intrinsic curvature of M proteins is essential for virus budding. While most current research has focused on the S protein, which is responsible for viral entry, and it has been motivated by the need to develop efficacious vaccines, the development of resistance through mutations in this crucial protein makes it essential to elucidate the details of the viral life cycle to identify other drug targets for future therapy. Our simulations will provide insight into the viral life cycle through the assembly of viral particles de novo and potentially identify therapeutic targets for future drug development.
}


\begin{document}

\section{Introduction}

Challenging every aspect of daily lives, the~severe acute respiratory syndrome coronavirus 2 (SARS-CoV-2), the~etiologic agent of COVID-19, has driven scientists all around the world to rapidly develop technology and research advancing our understanding of this virus and the means to combat it. The~coronavirus family extracts a significant burden on the global population through the sheer number of viruses from this family causing disease. Before~the pandemic, research had been hampered by the limited scope of severe infections dying out before becoming wide spread health risks (e.g., the original SARS-CoV) and by the high level of complexity involved in the formation and assembly pathways of these viruses. Up~to now, nearly all vaccines target the spike (S) proteins inhibiting viral attachment to the host cell membrane thus preventing viral entry into the host cell. Focusing on this single mechanism to disarm the virus can easily be rendered ineffective by mutations in the S~protein. 

\begin{figure}
    \centering
    \includegraphics[width=0.5\linewidth]{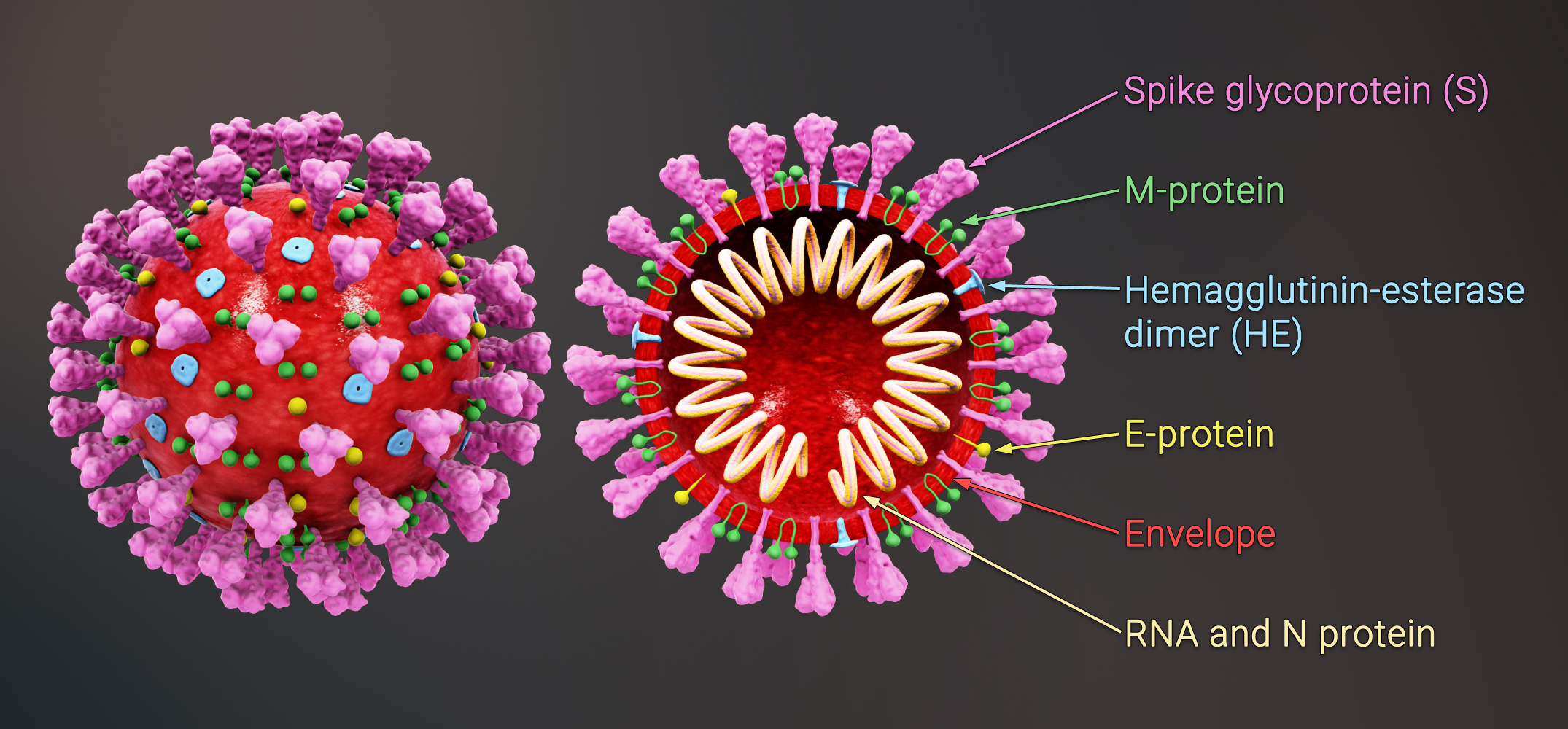}
    \caption{
    \hl{Schematic view of the structure of a coronavirus}. The~figure shows the structural proteins: Spike (S) proteins (\hl{magenta}), Membrane (M) proteins (\hl{green}) and Envelope (E) proteins (\hl{yellow}). The \hl{complex of genome and Nucleocapsid (N) proteins (beige) are} enclosed in a lipid membrane called an envelope (red).
    The picture is adapted from \url{https://www.scientificanimations.com/coronavirus-symptoms-and-prevention-explained-through-medical-animation}.} 
    \label{fig:corona}
\end{figure}

While the S protein does play a crucial role in the life-cycle of the virus, it is not integral to the assembly process. Coronaviruses have three other structural proteins: Envelope (E), Membrane (M), and~Nucleocapsid (N), which are known essential constituents for the assembly and formation of the viral envelope, \hl{see} Figure~\ref{fig:corona}. 
\hl{It is known that} the~N proteins \hl{attach to the} genomic RNA and form a compact viral ribonucleoprotein complex (RNP)~\cite{Chang2009,Chang2013,Chang2014,Klein2020,scientificamerican}. 
It appears that the electrostatic interaction between the positively charged N proteins and the negatively charged RNA is crucial for the genome condensation. Since coronaviruses have the largest genome ($\sim$30,000 nucleotides) among all RNA viruses, the~role of N proteins is specifically very important in condensing and packaging of viral RNA.  The~virus then forms as the compact RNP buds at the membrane of the Endoplasmic Reticulum-Golgi Intermediate Compartment (ERGIC) where other structural proteins (M, E, S) are located~\cite{Klein2020,Martinez-Menarguez1999,Klumperman1994}. 

The morphology of the formed RNP and how the N proteins interact with and condense the genomic RNA resulting in its packaging by a lipid envelope decorated with structural proteins inside a host cell are still unclear. Two models have been proposed based on Electron Microscopy (EM) images of RNP composite structures. Some studies~\cite{Chang2013,Chang2014,Macnaughton1978,Barcena2009} show that N proteins and RNA form a helical fiber structure (9--{16} {nm} in diameter), which attaches to the interior of the viral envelope. 
Other studies~\cite{Klein2020,Yao2020} suggest that the entire RNP consists of 35--40 copies of small but connected clusters in which each cluster contains 12 N proteins attached to 800 nucleotides forming a spherical layer with a thickness of 15~nm.

In addition to the uncertainty about the structure within the virus envelope, the~process of packing of RNP and its budding from the organelles inside the host cell is not well understood. In~fact, it is believed that the interaction between N and M proteins play an important role in the RNA encapsulation by the lipid envelope~\cite{Kuo2002,Kuo2016,Hurst2005,Huang2004,Arndt2010}.  Using cryo-EM, Neuman~et~al. have shown that the M protein forms a matrix which is responsible for the curvature of the viral envelope~\cite{Neuman2011,Neuman2020}. \hl{Several} other studies show that the presence of M proteins is required for virus budding and envelope formation~\cite{Mortola2004,Huang2004,Siu2008,Plescia2021}. It is worth mentioning that the M protein is the most abundant structural protein in the viral particle.  However, due to limitations of resolution in electron microscopy imaging, our knowledge remains rudimentary about the properties of all the structural proteins including M proteins composing the viral~envelope. 

In general, a~careful study of the assembly experiments reveals that there are very few known details about the formation and assembly of coronaviruses.  Computational models have often used to shed light on the myriad of complexities contributing to the assembly of other viruses not easily gleaned from experiments~\cite{Li2018,Panahandeh,Perlmutter2013,Chen:2007b,Fejer:10,Rapaport:04a,Patel2014,Vernizzi:2007a,Stefan}. All-atom molecular dynamics simulations provide powerful analytic tools in analyzing protein--protein interactions. However, because~these processes represent miniscule portions of an immensely larger enterprise and occur over time scales on the order of nanoseconds, they are nearly impossible to apply to the entire assembly and budding process~\cite{Collins2021,Yu2021,Cao2022}. To~this end, coarse-grained (CG) modeling has been employed to explore the kinetics of assembly of spherical icosahedral viruses. While a considerable number of CG computer simulations and the in~vitro self-assembly experiments of other viruses such as HBV~\cite{Ruan2018,Nair2018} have had remarkable impact on deciphering the factors that contribute to the assembly of viruses and the means to combat them~\cite{perlmutter2014pathways, Hagan2013, twarock2019structural, dykeman2014solving,Garmann2016,Garmann201909223,Chevreuil2018,Zandi2020,Comas-Garcia2014}, similar systematic studies aimed at understanding the assembly of coronaviruses are~missing.

To investigate and understand the assembly pathway of SARS-CoV-2 and gain insight into behavior over longer time scales, we have designed coarse-grained models for RNA, phospholipid membrane, and N and M proteins.  Using Molecular Dynamics simulations, we have examined what types of interactions between the membrane, proteins, and~genome are necessary to give rise to the packaging of RNA and the assembly of the virus.  Our simulations reveal the role of each structural protein in the formation of the SARS-CoV-2, the~mechanisms underlying the assembly process and the physical factors contributing to the virus morphology, see Video S6 and Video~S7.

We find that, while the positively charged N monomers can condense the negatively charged RNA, oligomerization of N proteins is essential for the packaging of the very long genome of coronaviruses. Without~N oligomerization, the~RNA will not be compact enough to be packaged. As~a result of the oligomerization of N proteins, RNA and N-proteins form several small helical clusters connected through small segments of RNA, which leads to the condensation of the long genome and its packaging. The~structures of RNP in our simulations reveal why some experimental results show that RNP has a helical structure~\cite{Chang2013,Chang2014,Macnaughton1978,Barcena2009}, while others find that it is a combination of several clusters with no particular order and symmetry~\cite{Klein2020,Yao2020}. In~fact, we find that it is a combination of the two observed cases. Our study shows that small helical clusters can join and condense RNA resulting in its packaging; however, one single long helical structure renders RNP too stiff to be~packaged.  
    
Furthermore, considering the length of RNA and the size of the virus, our study shows that the intrinsic curvature of M proteins is necessary for viral budding. Under~no circumstances were we able to observe the packaging and formation of an envelope through budding in the absence of the M protein intrinsic curvature. Consistent with the experimental data on most coronaviruses~\cite{Neuman2020,Siu2008,Plescia2021}, we find that, while the interaction between the M and N proteins is not absolutely necessary for the budding process of an empty envelope, it is necessary for the genome packaging and significantly facilitates the virus budding through the host cell~membrane.

Our work elucidates the fundamental process of viral formation essential to its reproduction, survival and ultimately its infectivity.  Even though our knowledge about the coronavirus proteins responsible for the virus assembly is still very basic, our CG model based on the known protein structures has been able to successfully condense the RNA and shed light on the kinetic pathways of the viral assembly. Our simulations have deciphered to some extent the elusive assembly process of SARS-CoV-2, which could be the basis and guide for the design of future~experiments. 

Understand that the factors contributing to the formation of coronaviruses can ultimately prevent viral replication through disruption of the assembly. It is through inhibition of multiple viral mechanisms that resistance is most effectively thwarted and viral replication~suppressed.  

\section{Materials and~Methods}

\subsection{N~Protein}
We construct our coarse-grained model for the N proteins based on the fact that they are highly positively charged and mainly form dimers in solution.
Furthermore, NMR, small angle X-ray scattering, and other microscopy experiments~\cite{Zeng2020,Yang2021,McBride2014} reveal that the structure of N proteins (built from 419 amino acids) can be divided into five regions, namely: the N-terminal disordered region (N$_\text{IDR}$), the~RNA binding domain (RBD), the~Linker disordered region (Linker$_\text{IDR}$), the~C-terminal dimerization domain (CTD), and the C-terminal disordered region (C$_\text{IDR}$), in~which N$_\text{IDR}$, Linker$_\text{IDR}$ and C$_\text{IDR}$ form the highly flexible intrinsically disorder regions (IDRs) of the protein~\cite{Cubuk2021,Perdikari2020}. 
The in~vitro experiments using reconstitution approaches and cellular assays have shown that the Linker region plays a crucial role in phase separation of the N protein with RNA~\cite{Lu2021}, whereas the CTD mainly contribute to the N protein oligomerization~\cite{Chang2014}.

\begin{figure}
    \centering
    \includegraphics[width=0.7\linewidth]{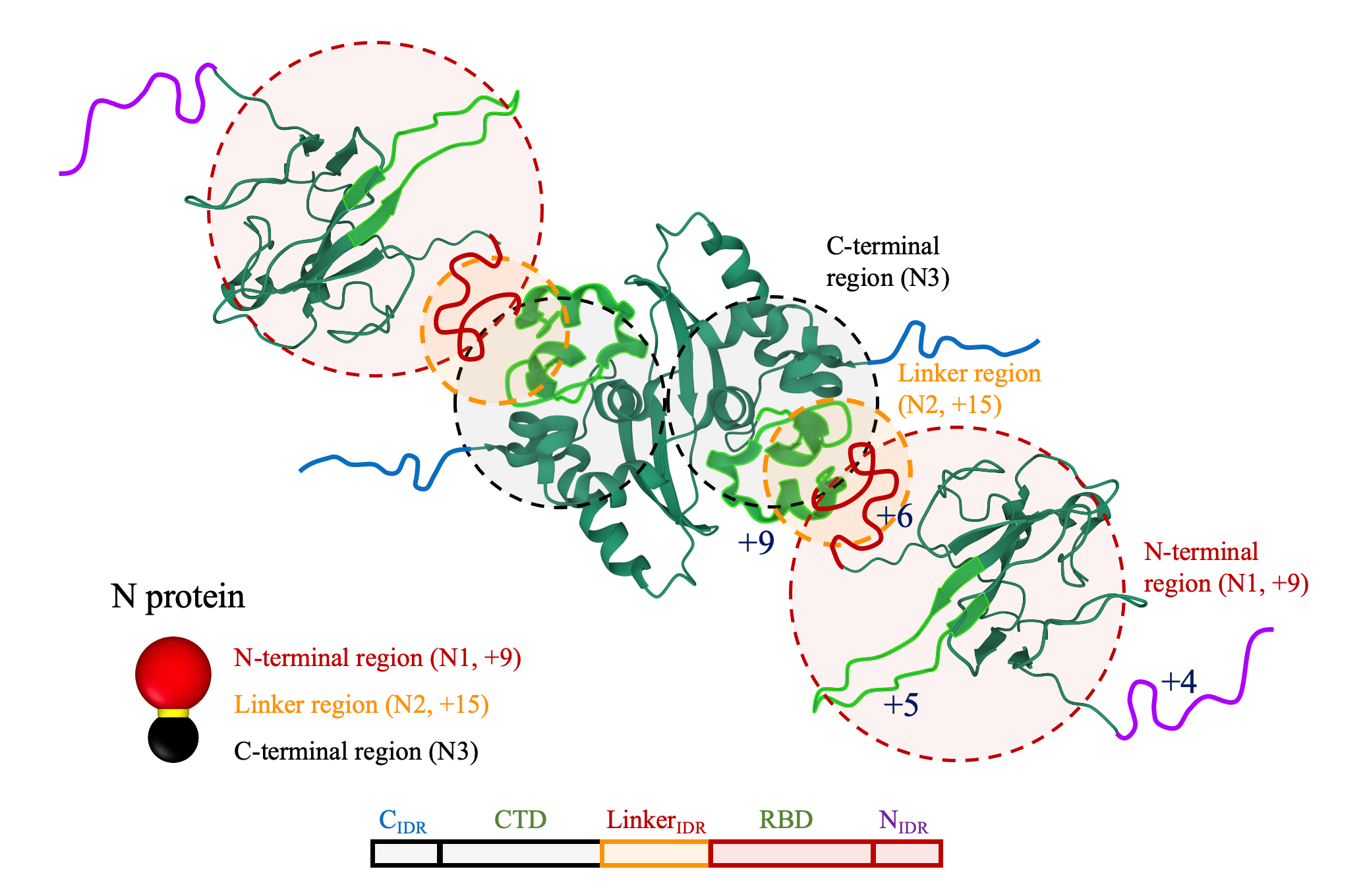}
    \caption{Schematic view of an N protein. The~dimerization domain (CTD) of the N protein and the RNA binding domain (RBD) are obtained from PDB 6ZCO and 6YI3, \hl{respectively}~\cite{Zinzula2021,Dinesh2020}. 
    The~intrinsic disorder regions (C$_\text{IDR}$, Linker$_\text{IDR}$, N$_\text{IDR}$) are obtained from UniProt P0DTC9. The~visualization is performed through RCSB 3D-view, where the concentrated positive charges are colored light green; the rest of the regions have a zero net charge. The~shaded circles indicate the coarse-grained model including three spherical particles: the N-terminal region (N1, red), the~linker region (N2, yellow), and the C-terminal region (N3, black).}
    \label{fig:modelN}
\end{figure}

A careful review of the entire 419 residues of N proteins show that the majority of basic residues are located in N$_\text{IDR}$ (+4, residues 1--43), the~middle of RBD (+5, residues 89--110), Linker$_\text{IDR}$ (+6, residues 181--243), and~the N-terminal of the CTD (+9, residues 244--278), as~shown in Figure~\ref{fig:modelN}. Since our focus is on the electrostatic interaction between the N proteins and RNA, we assume that the N protein is composed of three regions based on its charge distribution. A~schematic view of our N protein model is shown in Figure~\ref{fig:modelN}.  Each N protein consists of three hard particles: N-terminal (N1, red), Linker (N2, yellow), and C-terminal (N3, black) regions, respectively. Note that the C-terminal region contains C$_\text{IDR}$ and CTD, while the N-terminal region contains N$_\text{IDR}$ and RBD. We note that,
since the charges in CTD are located at the N-terminal sites, we combine them with the Linker$_\text{IDR}$ charges, which form the N2 region with +15 charges, the~yellow particle in Figure~\ref{fig:modelN}. 

We further perform a set of molecular dynamics simulations to adjust the strength of N--N interaction based on the experimental data reporting that most N proteins form dimers in solution~\cite{Zeng2020}.
We find that the oligomerization of N proteins largely depends on the relative size of N1, N2, and N3 domains as shown in Figure S1. This is mainly due to the fact that a smaller N3 effectively increases the electrostatic repulsion between charged regions (N1 and N2) rendering the higher order oligomerization (trimers, tetramers, \dots) less favorable.
Based on the analysis of the distributions of the N protein oligomerization, we set the diameters of N1, N2, and N3 particles equal to $1.2a$, $0.5a$, and $0.8a$, respectively, and set $\epsilon_{N_3N_3}=20$ for the N3{--}
N3 interaction, where $a$ is the system unit length corresponding to 3 nm. With~these parameters, the~majority of oligomers form dimers regardless of the N protein concentration, consistent with the experimental~data.
\subsection{Membrane and Membrane~Proteins}
\begin{figure}
    \centering
    \includegraphics[width=\linewidth]{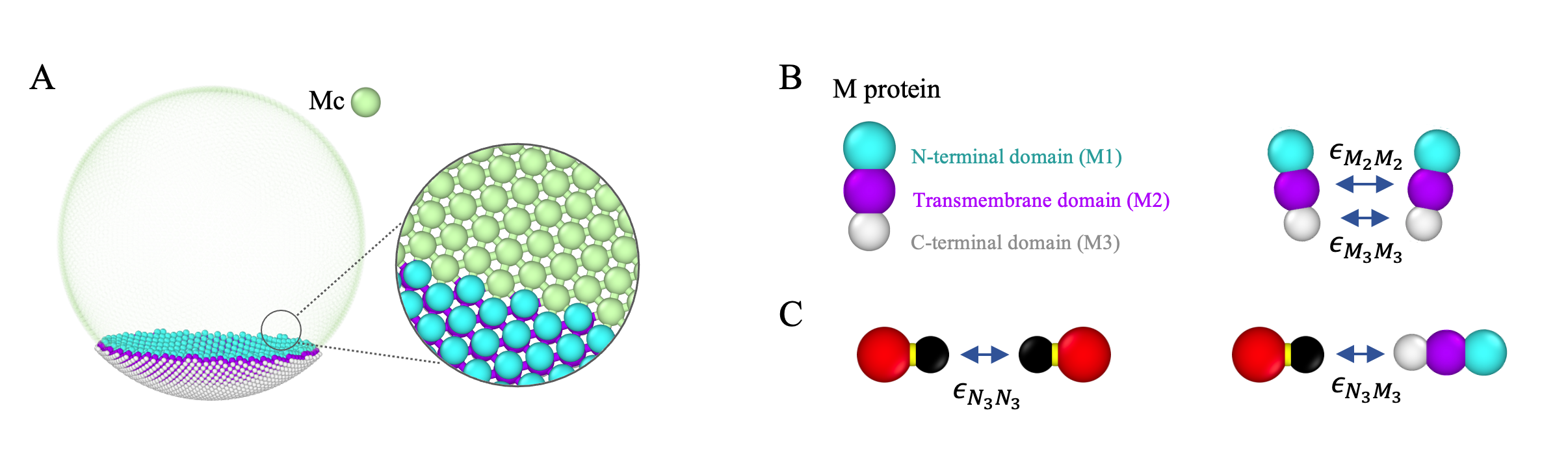}
    \caption{(\textbf{A}) A~schematic view of ERGIC. The~membrane vertices (Mc) are green and the M protein particles are colored blue-purple-white. The~right figure shows the bond connections in the ERGIC membrane built from a triangular lattice.  (\textbf{B}) a~schematic view of an M protein. The~coarse-grained M protein model contains three hard particles: M1 (N-terminal domain, blue); M2 (Transmembrane domain, purple), and M3 (C-terminal or cytosolic domain, white). The~arrows indicate the attractive M2--M2 and M3--M3 interactions. The~diameters of M1 and M2 are equal while that of M3 is smaller than the other two, which is the source of the intrinsic curvature of M proteins; (\textbf{C}) illustration of the N--N (left) and N--M (right) interactions. The~attractive interactions between domains are indicated with arrows, where the C-terminal of the N proteins (N3) attracts each other, and N3 also attracts the C-terminal of M proteins (M3).}
    \label{fig:modelM}
\end{figure}

We model the ERGIC membrane as a spherical vesicle built of a triangular lattice; see Figure~\ref{fig:modelM}A. The~energy of the triangular network involves both the bond stretching and bending energies.  The~stretching energy can simply be expressed by a harmonic potential summed over all bonds as follows:
\begin{eqnarray}
    E_s /k_BT &=& \sum_{<bond>} \frac{1}{2} k_s (l - l_0)^2, 
\end{eqnarray}
where $l_0$ is equal to $a$, the~system length scale corresponding to 3 nm as described before, and~$k_s=20a^{-2}$ is the stretching modulus.
The bending energy is determined by summing over all pairs of triangular subunits sharing an edge and can be written as
\begin{eqnarray}
    E_b /k_BT&=& \sum_{<bond>} k_b (1 - \cos(\theta-\theta_0)),
\end{eqnarray}
with $\theta_0=\pi$ the preferred angle and $k_b=20$ the bending modulus.
To make the membrane incompressible, we have placed a spherical hard particle with diameter $a$ (green sphere in Figure~\ref{fig:modelM}A) on each membrane~vertex. 

As for the M protein, we have modeled it as a rigid body composed of three particles, M1, M2, and M3, corresponding to N-terminal, transmembrane, and C-terminal domains, respectively, see Figure~\ref{fig:modelM}B. We have then replaced a fraction $f$ of the membrane vertices with membrane proteins as illustrated in Figure~\ref{fig:modelM}A.

The M2--M2, M3--M3, N3--N3, and N3--M3 (see Figure~\ref{fig:modelM}B,C) interactions are modeled by Lennard--Jones potential,
\begin{align}
LJ_{ij} /k_BT&= \sum_{i}4\epsilon_{ij}((\frac{\sigma}{r_{ij}})^{12} - (\frac{\sigma}{r_{ij}})^6),
\end{align}
where $\sigma=2^{-1/6}d$ with $d$ the distance between two interacting particles when touching each other and $\epsilon_{ij}$ and $r_{ij}$ are the interaction strength and distance between two particles $i$ and $j$, respectively. The~interactions are subject to a cutoff distance $r_{cut} = 3a$. 

The electrostatic interactions between N2, N3, and the genome are modeled by Debye--Hückle interaction potential,
\begin{align}
DH_{ij}/k_BT &= l_{b} Z_i Z_j* \frac{e^{\kappa d}}{ 1 + \kappa d} \frac{e^{-\kappa r_{ij}}}{r_{ij}},
\end{align}
where the Bjerrum length $l_{b}=0.24a$ corresponds to the length scale over which the electrostatic interaction between two elementary charges is comparable to the thermal energy, $Z_i$, and $Z_j$ are the charge numbers, $d$ is the sum of the radius of two particles, and~$\kappa^{-1}=0.5a$ is the Debye–Hückel screening length.
In addition, the~excluded volume interaction between all particles is presented with a shifted LJ repulsion potential
\begin{equation}
LJ^{rep}_{ij}/k_BT = \sum_{<i,j>}4\epsilon((\frac{\sigma}{r_{ij}})^{12} - (\frac{\sigma}{r_{ij}})^6) + \epsilon,
\end{equation}
with $\epsilon=1$, $r_{cut} = d$ and $\sigma = 2^{-1/6}d$.

\subsection{Genome}
\revSL{While the length of SARS-CoV-2 RNA is 30k nucleotides, it is important to note that 63.8\% of the RNA is base-paired~\cite{Cao2021}.  Given that dsRNA forms an A-form helix which is 20--25\% shorter than dsDNA helix~\cite{Lipferta2014, Arias-Gonzalez2014}, we assume each bead is built of 11--12 bp. Thus, we use 800 beads to present the double-stranded segments of RNA. We note that the single-stranded segments of RNA are mostly at the branching points or stem loops~\cite{Klein2020}.  For~simplicity, we did not explicitly model the single-stranded segments but rather implicitly consider their role in RNA flexibility. To~this end, we model RNA as a negatively charged linear chain composed of L = 800 beads whose diameter is 1.0a and are connected by unextendable bonds. The~chain is the so-called freely-jointed model if we do not consider the electrostatic repulsion.}
Because of the base-pairing and the secondary structures, RNA has often been modeled as a branched polymers~\cite{Perlmutter2013,Li2017a,Nguyen-Bruinsma}.  Thus, we also consider the case of branched polymers with the same total length $L=800$.  

Taking into account the condensation of counter ions around the genome~\cite{Manning1978}, we set the effective number of charges of each coarse-grained genome bead equal to $Z_g=\frac{b}{z l_b}Q_g$ with $b$ the distance between the genome phosphate charges, $l_b$ the Bjerrum length as described above, $z$ the valence of the counter ions, and~$Q_g$ the original charge number of each bead.  In~the following sections, we perform a number of simulations for $Z_g\in[-2, -10]$ and employ the Debye--Hückle theory to model the protein--RNA electrostatic~interactions.

\subsection{Simulations}
\begin{figure}
    \centering
    \includegraphics[width=0.5\linewidth]{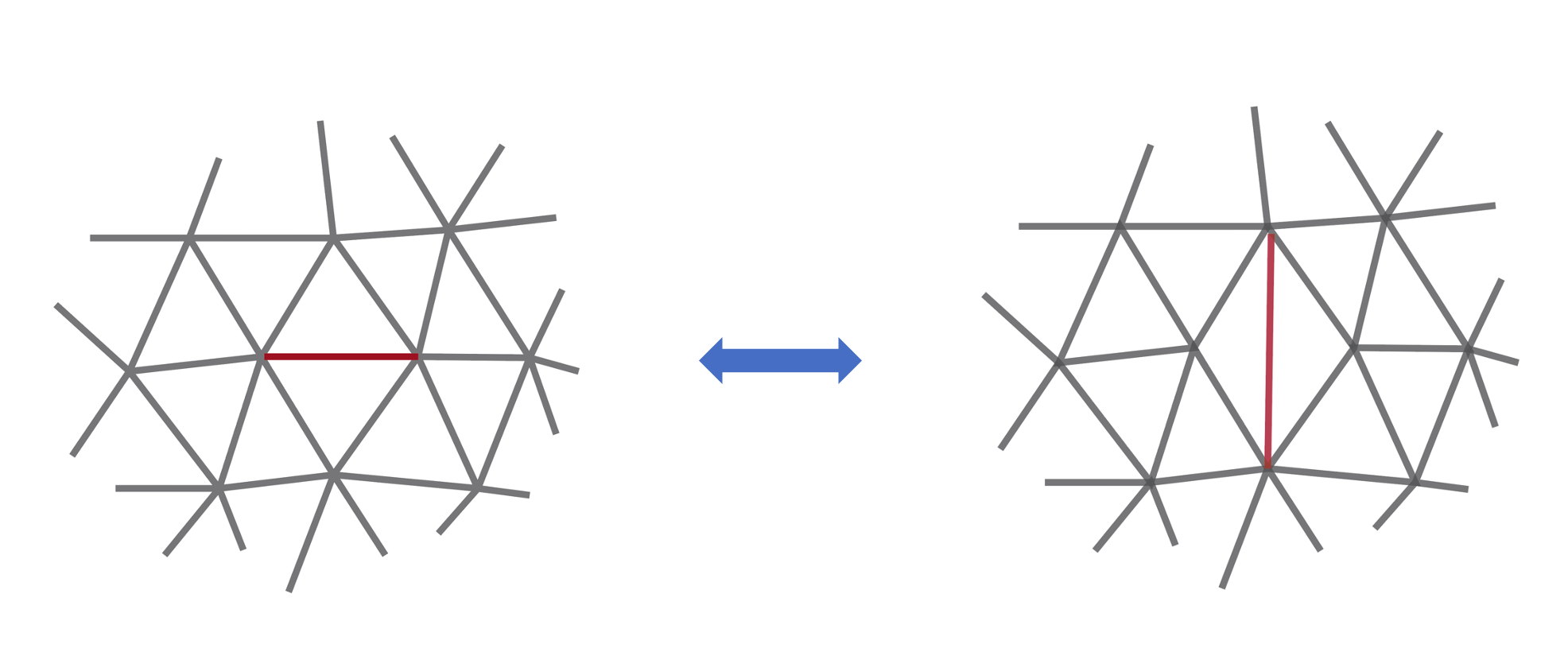}
    \caption{A schematic diagram of the Bond-Flip Monte Carlo method in which the shared edge of two adjacent triangles are clipped and reconnected to the diagonal vertices to change the local topology of the \hl{structure}~\cite{panahandeh2018equilibrium}.} 
    \label{fig:edgeflip}
\end{figure}

To study the kinetics of budding, we use the Lagenvin integrator to integrate the membrane vertices movement, with~time step $dt = 10^{-3}s$.
At every 100 time steps, we perform Monte Carlo (MC) simulations combined with the bond flipping method to change the membrane connectivity.
The details of Bond Flip moves are shown in Figure~\ref{fig:edgeflip}.  Each MC step consists of $N_b$ attempts of removing a randomly chosen bond between two triangles and attaching two vertices that were not linked before as shown in Figure~\ref{fig:edgeflip}, where $N_b$ is the total number of bonds. Detachment and attachment of the bonds are such that the total number of vertices and bonds remain constant.
The simulation is performed on {NVIDIA GeForce RTX 3090} 
where a 8000 s run takes approximately 3 h. We have employed HOOMD-blue~\cite{hoomd} for the Molecular Dynamics (MD) simulations and \hl{built a plug-in} 
for Bond-Flip MC~simulations. \hl{All simulations are visualized by OVITO software}~\cite{ovito}.

\section{Results}
\subsection{N Protein and RNA Assemble into RNP with Multiple Protein Clusters}
As noted in the Introduction, coronaviruses have the largest genome among all single-stranded RNA viruses. Several experiments have shown that the presence of N proteins is necessary for the packaging of such a long genome by the viral envelope. In~fact, it is now widely accepted that the electrostatic interaction between the positively charged N proteins and negatively charged RNA is the driving force for genome~encapsulation. 

\begin{figure}
\centering
\includegraphics[width=\linewidth]{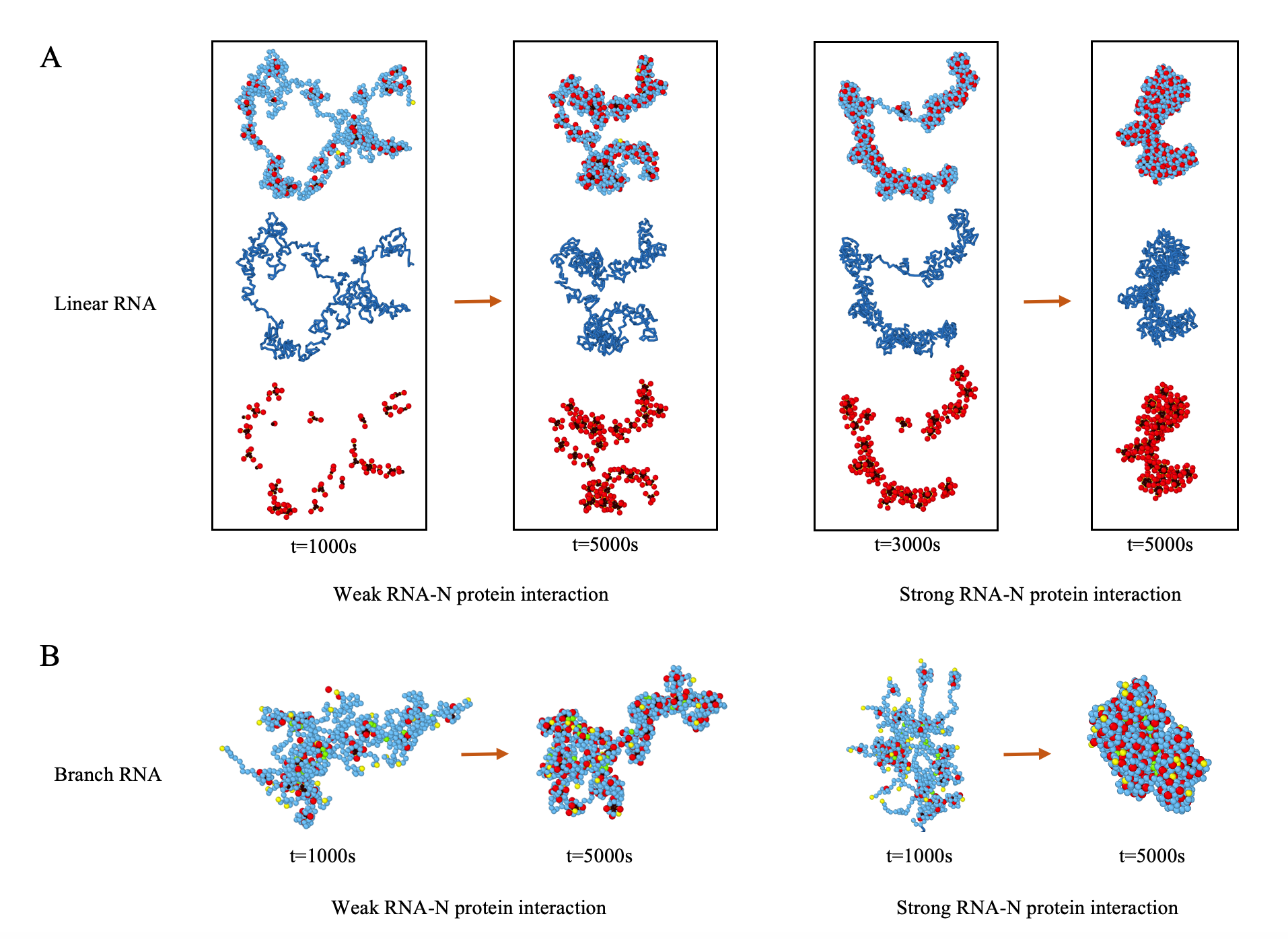}
\caption{(\textbf{A}) N proteins condense a linear chain with $L=800$ subunits and form RNP. Both cases of the weak ($Z_g=-2$) and strong ($Z_g=-5$) RNA--N protein interactions are presented.  The~N protein concentration is 25$\si{\micro M}$. 
In each window, the~upper snapshot shows the RNP with genome in blue and N proteins in red and black. The~middle and lower snapshots display the RNA and N protein clusters (NPCs), respectively.  Note that, if the distance between the N-terminals of N proteins (black) are less than a cut-off distance, they are considered to belong to the same cluster. 
(\textbf{B}) N proteins condense RNA.  RNA is modeled as a branched polymer with the branch number $N=30$.  Both cases of the weak and strong RNA--N protein interactions are considered.  The~branched points and end points are colored in green and yellow, respectively. At~the weak RNA--N protein interaction, the~N proteins aggregate into clusters in the same way as they do in the case of linear chains, regardless of the branching structure of the chain. The~situation differs for the case of the strong RNA--N protein interaction, where the N proteins condense along the branches first, which later aggregate to form a compact RNP.}
\label{fig:RNP}
\end{figure}

In this section, we explore the condensation of RNA by N proteins into a compact ribonucleocapsid structure (RNP) as a function of the concentration of N proteins and the strength of genome--N protein interaction.  Following the literature on the structure of the N protein, we have modeled it as described in the Methods section; see Figure~\ref{fig:modelN}. Note that we have chosen the N--N interaction based on the structure shown in Figure S1, \revSL{where the dimers are dominant structures in the solution. It is worth mentioning that the stronger N--N interaction results in the formation of higher-ordered oligomers, but, as it is illustrated in Figure S3, they do not modify the final structure of RNP.} Using these subunits, in~the majority of cases, we find that N proteins form several clusters along the length of RNA.  Quite interestingly, as~illustrated in Figure~\ref{fig:RNP}, RNA wraps around N proteins to form several clusters with helical structures that are connected by small segments of naked~RNA. 

To quantify the structure of RNPs and to create a basis for the comparison with the future experiments, we have analyzed the clusters and assigned a unique index number and color to each one, see Figure S2. If~the distance between two neighboring N proteins is less than a cutoff distance $1.2a$ with $a$ the system length scale {3} {nm}, then we consider that they belong to the same~cluster.

Figure~\ref{fig:RNP}A shows the snapshots of the simulations of the formation of several clusters built from N proteins along a linear genome for two cases of weak and strong RNA--N protein interactions. The~top row illustrates the RNA--N protein (RNP) complex structures. For~clarity, the~middle row displays RNA only, whereas the bottom row illustrates only the N protein clusters (NPC) without showing the genome.
The simulations indicate that multiple nucleation sites simultaneously form along the genome. At~the early stages of the assembly, we observe that RNA wraps around N proteins (see $t$ = 1000 s in Figure~\ref{fig:RNP}A) and several small clusters made up of two or three N proteins are formed. As~time passes, the~small clusters either fuse with neighboring clusters or attract more proteins from the solution to form larger~clusters. 

We note that the strength of RNA--N protein interaction in the simulation depends on the effective genome charge density, which depends on the concentration of salt and counterions in solution~\cite{Belyi2006}. 
As the effective charge density increases, both the RNA--N protein attractive interaction and the RNA self-repulsion become stronger. At~this regime, larger N protein clusters form with an average of 10 proteins in each cluster. As~time passes, multiple clusters join and form a large cylindrical configuration which is pretty rigid, as~shown on the right side of Figure~\ref{fig:RNP}A. 

\revSL{We recall that the focus of our work is on the regimes in which electrostatics dominates all the other interactions between RNA and N proteins.  If,~however, the~strength of RNA--N protein interaction were weak, the~presence of packaging signals or other specific interactions would be necessary to produce a similar genome condensation process~\cite{deBruijn2022}. In~that case, each packaging signal would act as a nucleation site for the N proteins to start forming clusters along RNA.}

The snapshots of simulations of the formation of RNPs using branched polymers are shown in Figure~\ref{fig:RNP}B.  In~each case, $30$ branch points were randomly generated. The~length of each branch was randomly chosen between one and forty monomers while the total length of the branch polymer was kept the same as the linear chain model ($L=800$). Our simulations show that, at the weak RNA--N protein interaction, several clusters immediately nucleate along the longest length of the polymer, which later will grow in all directions. However, as~we increase the strength of RNA--N protein interaction (the effective genome charge density), the~N proteins nucleate first along the small branches and then they grow towards the branch point, see Video~S1.  

Moreover, while the branchiness of the polymers seems to speed up the condensation process, we find that the impact of the secondary structure of RNA becomes more significant when the branch number is larger than $10$. At~the smaller branch numbers (\hl{$\le$}10), the~condensation of branch polymers follows more or less a similar pathway as a linear chain (see Video S2). \revSL{We note that, while previous experimental and theoretical investigations on small icosahedral viruses like CCMV and BMV have shown that branchiness increases packaging efficiency and lowers the free energy of the formation of virus particles~\cite{Comas-Garcia2012,Perlmutter2013,Li2017a}, the~effect of branchiness on the rate and the kinetics of virus assembly has not yet been studied. }

We have also explored the effect of the protein concentration on the structure of RNP (see Videos S3 and S4). 
The results of the simulations indicate that the protein concentration does not have a significant impact on the final RNP structures if the genome effective charge density is low. Nevertheless, a~higher protein concentration speeds up the condensation time as expected.  For~the case of the genomes with a higher effective charge density, a~higher protein concentration, however, facilitates the adhesion of neighboring clusters, rigidifies RNP configuration, and~increases the number of clusters and the number of N proteins per cluster, leading to a more compact RNP~configuration.

Finally, we examine the impact of the strength of the attractive hydrophobic interaction between N proteins on the RNP structure.  We find that the N--N attractive interaction plays a key role in the formation of N protein clusters along RNA. Even a weak interaction between N proteins can induce the formation of several small clusters.  In~fact, for~a wide range of the strength of the N--N interaction, the~RNP configurations are almost the same, and the number of N proteins in clusters remains mainly the same (see Figure S3). 
Quite interestingly, we find that these clusters are crucial for making RNP compact enough to ensure its packaging.  In~the absence of the N--N interaction, the~positive charges on N proteins condense the negatively charged RNA but not as much as it is necessary for its complete packaging.  The~presence of N protein clusters along the genome is essential for RNP packaging if the genome is very long as in the case of SARS-CoV-2, see the structures of RNP in Figure S3 for $\epsilon_{N_3N_3}=0$ and $\epsilon_{N_3N_3}=30$.

\subsection{Spontaneous Radius and Line Tension of M Proteins Are Important for Budding}

As noted in the Introduction, despite the intense experimental and theoretical studies, the~most basic questions about the mechanism and relevant time scales associated with the budding process of coronaviruses are still lingering. 
Using a combination of electrophoresis, fluorescence and electron microscopy, the~earlier experiments on coronaviruses have revealed that M, E, and N proteins are sufficient for the budding of virus-like particles (VLPs), whereas the shape of VLPs is mainly determined by M proteins~\cite{Mortola2004,Huang2004,DeDiego2007,Siu2008,Neuman2011,Xu2020,Plescia2021}. It is important to note that, according to the previous experiments, the~E proteins are required for budding; it appears that they contribute to the completion of assembly and the scission of the budding neck.
In the absence of enough information about the role of E proteins, we do not explicitly include their effects here, and~thus investigate the general properties of the proteins embedded in the lipid membrane resulting in virus budding.
More specifically, we investigate the minimum requirement for the packaging and budding of RNP at the ERGIC membrane and explore the contribution of the mechanical properties of M proteins to the formation of viral~envelope. 

As explained in the Methods section, we model ERGIC as a fluid membrane and model M proteins based on the literature investigating its structure~\cite{Cao2022,Thomas2020,Heo2020,Dolan2022}. To~explore the impact of M--M interaction on the formation of the envelope, we first build a two-dimensional phase diagram as a function of the interaction between the transmembrane domains (M2--M2) and endodomains or cytosolic domains (M3--M3) of M proteins, see Figure~\ref{fig:modelM}. We note that, for all simulations presented in this section, we have assumed that M proteins have already aggregated into a circular region and that ERGIC is relaxed with several pentameric defects located at various positions. It is worth mentioning that we have explored different scenarios and found that the locations and the initial numbers of defects do not affect the final budding configuration. We explain in the later section why the initial distribution of M proteins in the membrane does not have an impact on the final~results.

\begin{figure}
\centering
\includegraphics[width=\linewidth]{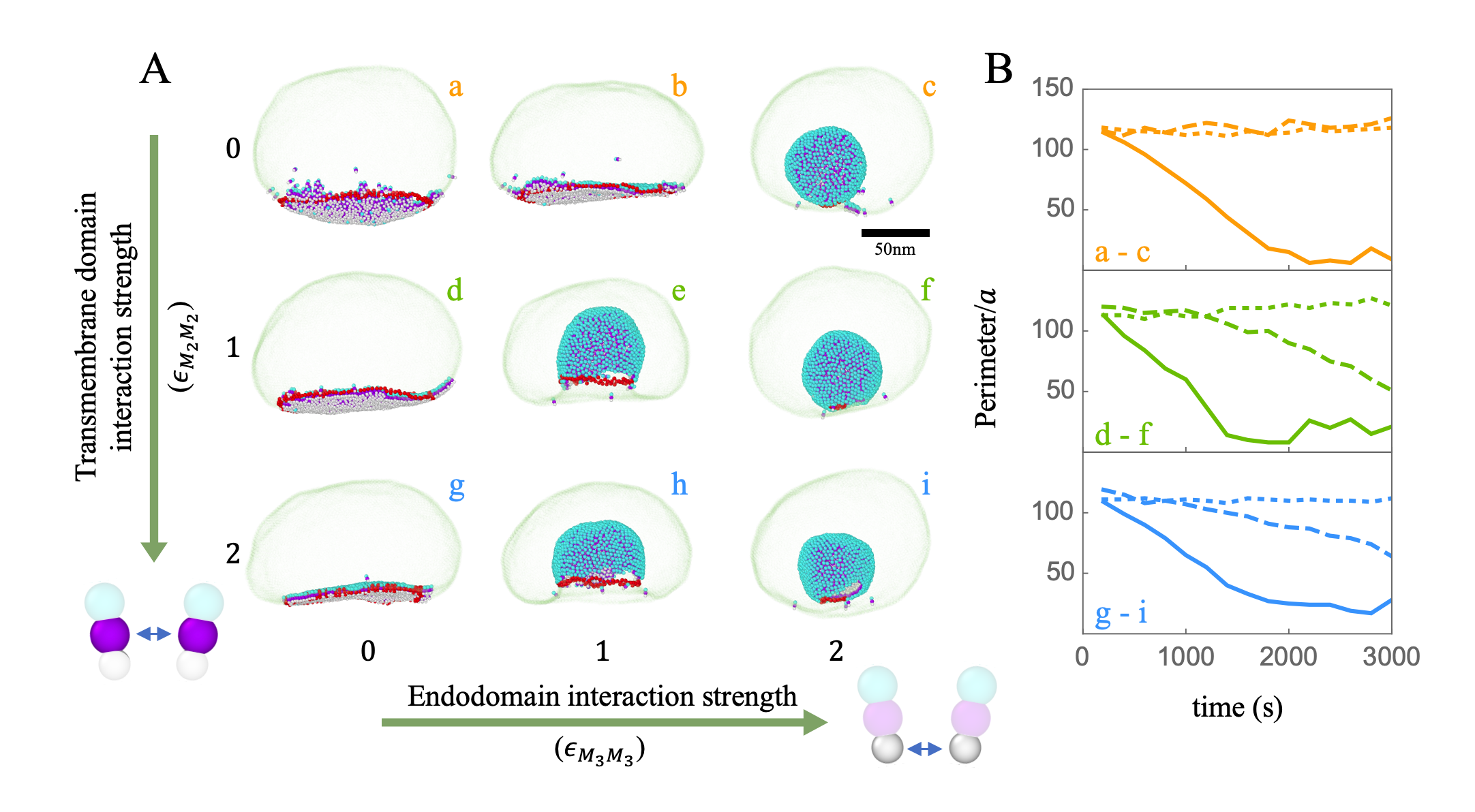}
\caption{(\textbf{A}) {Snapshots} 
 of budding through the ERGIC membrane as a function of the strength of the interaction between the M protein transmembrane domains ($y$-axis) and between the endodomains ($x$-axis). The~interactions become stronger from top to bottom and from left to right. The~lipid--lipid interaction strength is kept constant $\epsilon_{McMc}=1$ for all simulations;
(\textbf{B}) the~perimeters (in the unit of $a$) of the cap built from the M proteins as a function of time in the budding process. The~dotted, dashed, and solid lines correspond to the endodomain interaction strengths $\epsilon_{M_3M_3}$ = $0$ \hl{(a, d, g)}, $1$ \hl{(b, e, h)} and $2$ \hl{(c, f, i)}, respectively. As~the strength of $\epsilon_{M_3M_3}$ becomes stronger (the solid line), the~budding process becomes faster.}
\label{fig:ERGIC}
\end{figure}

Figure~\ref{fig:ERGIC}A shows the snapshots of the M proteins budding at ERGIC in the absence of a genome. The~strength of the interaction between the segments of the proteins embedded in the membrane (M2--M2) increases from top to bottom while the strength of interaction between cytosolic domains, the~segments outside of the membrane (M3--M3), increases from left to right.  Figure~\ref{fig:ERGIC}B shows the plots of the size of the perimeters of the partially budding vesicles (red regions in Figure~\ref{fig:ERGIC}A) vs.~time. 
The snapshots and time analysis of the growing vesicles indicate that the stronger transmembrane-transmemebrane (M2-M2) and endodomain-endodomain (M3--M3) interactions facilitate the budding process within the regime explored in Figure~\ref{fig:ERGIC}A. 
We emphasize that we have set the size of endodomains smaller than the transmembrane domains, which gives rise to the M protein's effective spontaneous curvature, leading to the bending of the membrane. A~stronger attractive interaction between endodomains facilitates budding and overcoming the energy barrier for the formation of~vesicles. 

It is worth mentioning that, since the transmembrane domains are embedded in the membrane, their interactions with each other and the membrane create a line tension, promoting the budding process through reducing the perimeter of the budding region. However, if~the strength of the transmembrane-transmembrane interaction is beyond a certain limit, the~M proteins can easily get trapped in a metastable state where the proteins form a hexagonal lattice and cannot bud. Thus, we were not able to locate a regime in which the line tension by itself is sufficient for virus budding, see the first column in Figure~\ref{fig:ERGIC}A. The~spontaneous curvature of M proteins is essential for bending the membrane to complete the budding and viral formation.  It is interesting to note that, even in the absence of the M2--M2 attractive interaction, if~the strength of M3--M3 attractive interaction is high enough, the~budding will take place (see Figure~\ref{fig:ERGIC}A.c).

We note that recent computational models on SARS-CoV-2 show that M proteins can induce the membrane curvature through the protein--lipid interaction~\cite{Collins2021}, which can be equivalent to the M3--M3 (cytosolic domains) interaction in our model. However, based on the fact that the formation of empty envelope is not optimal for a virus, we think that the effective M3--M3 interaction of SARS-CoV-2 is not sufficient for budding and that the RNP is required to trigger the budding so that the genomic material can be packaged more efficiently. 
In fact, some experiments on coronaviruses suggest that RNP interact with M proteins through the electrostatic interaction between the negatively charged portion of the C-terminal domain of the N protein and the positively charged cytosolic domain of the M protein~\cite{Kuo2002,Kuo2016b,Verma2006,Luo2006}. In~the next section, we show how N proteins can induce viral budding even if M3--M3 interaction is~weak.

\subsection{Two Different Models for RNA Packaging}

\begin{figure}
    \centering
    \includegraphics[width=\linewidth]{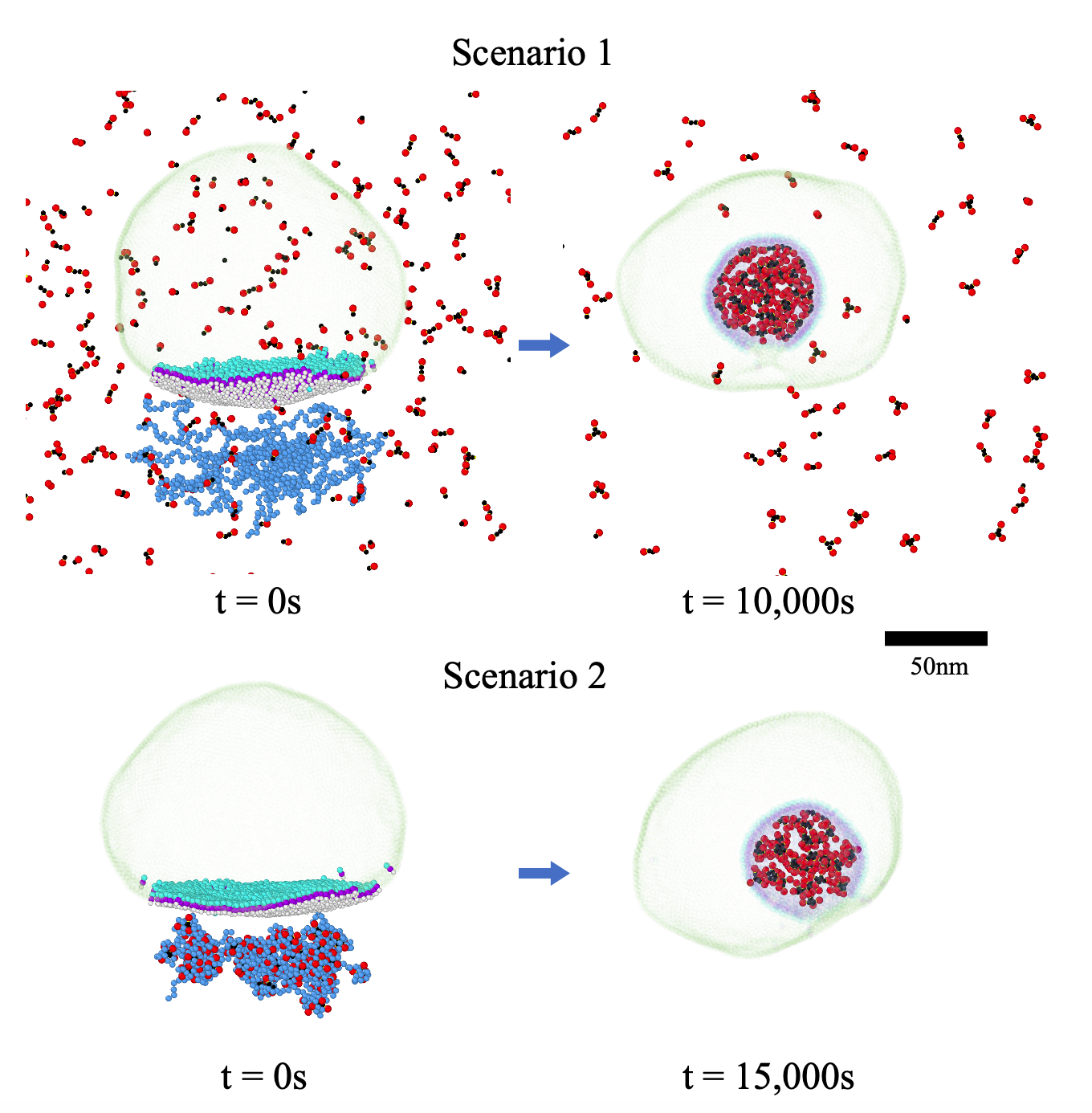}
    \caption{{Snapshots} 
 of simulations of the budding process for two different cases: Scenario 1. Condensation of the genome occurs at the same time as the budding process. In~this case, at~the beginning of the simulations, the~naked branched genome is located near the M proteins embedded in the ERGIC membrane surrounded by N proteins (25 $\si{\micro M}$), where the N-terminals of N proteins attract the genome and their C-terminals interact with M proteins, triggering both RNP condensation and budding simultaneously.
    Scenario 2: N proteins (25 $\si{\micro M}$) have already condensed RNA before the transport of RNP to the ERGIC interface. In~this scenario, the~N proteins within RNP are arranged such that their C-terminals interact with the endodomain of M proteins and trigger the budding process. For~clarity, only the N proteins (not RNA) are shown after the completion of the budding process in both cases. 
    For both scenarios, the~genome length is $L=800$ and genome charge is $Z_g=-2$, the~interactions between proteins are $\epsilon_{McMc}=1$, $\epsilon_{M_2M_2}=1$, $\epsilon_{M_3M_3}=0$, $\epsilon_{N_3N_3}=20$, and~$\epsilon_{N_3M_3}=10$.}
    \label{fig:scenario}

\end{figure}

We now explore the effect of the interaction between N and M proteins on the packaging of RNP at the ERGIC membrane.
We consider two different scenarios: (1) The genome has not yet been condensed but is surrounded by N proteins and is located in the vicinity of M proteins embedded in ERGIC (Video S6), and~(2) the compact RNP has been formed and transported to the ERGIC interface before interacting with the M proteins (Video S7).  The~snapshots of the simulations of two scenarios are shown in Figure~\ref{fig:scenario}. 

A careful review of the snapshots shows that both scenarios are feasible and could result in the formation of viral particles. The~cluster analysis of the final structures (Figure~S5) reveals that, when the genome is not condensed but is in the vicinity of the membrane surrounded by N proteins, many small RNP clusters form
right at the beginning of the assembly process. This is mainly due to the fact that the interaction with M proteins make the small clusters consisting of two to four N proteins survive over a relatively longer period of time. The~packaged RNP through this mechanism contains more N proteins than if it were formed while away from the membrane and then would be transferred to the membrane to interact with the M proteins to acquire its envelope. In~the latter case (Scenario 2), the~number of clusters remains constant during the budding as the clusters have already been formed and fully relaxed. However, the~packaging of RNP will take more time as N proteins in the clusters need to reorient and expose their C-terminal domains to interact with the M proteins. \revSL{It is worth mentioning that the protein concentration does not affect the rates of condensation and the assembly process, in~the case of the second scenario. However, for~the first scenario, the condensation and budding occur simultaneously so that the protein concentration and thus rate of condensation will have an impact on the budding rate.}

Furthermore, we find that the strength of the attractive interaction between the N proteins and RNA modifies the structure of RNP inside the envelope. The~clusters along RNP are distributed more sparsely inside the viral particle for a weak N protein--RNA interaction while RNP assumes a more condensed structure if the N protein--RNA interaction is stronger, as~shown in Figure S6.  In~the case of the second scenario, such condensed conformation might prevent RNP from being packaged at the ERGIC membrane since the N protein clusters are too compact to reorient their C-terminal and interact with M proteins. A~thorough analysis of RNP structures can reveal the condensation pathway of the coronavirus genomic RNA. However, the~current cryo-em images cannot distinguish the difference between the structures of RNP formed through different pathways. The~comparison of our simulations with future cryo-em images and experiments with different mutants of N proteins can shed light on the mechanism through which RNP is condensed and packaged at~ERGIC.

Our simulations also show that, if we set the interaction between transdomains $\epsilon_{M_2M_2}=1$ but between endodomains $\epsilon_{M_3M_3}=0$ (see Figure~\ref{fig:modelM}), the~M--M interaction cannot bend the membrane and no empty vesicle can form, as~shown in Figure~\ref{fig:ERGIC}Ad.  However, in~the presence of RNP,  the~interaction between M and RNP is enough for the virus to bud even though $\epsilon_{M_3M_3}=0$. We emphasize that, if the size of M3 is not smaller than M2, {under~no circumstances does the virus bud as discussed in the previous section, indicating that the spontaneous curvature of M proteins is absolutely necessary for budding} 
(see Figure S4).  It appears that M--N interaction constitutes the main contribution to the RNP budding if there is no interaction between endodomains.  Because~of the attraction between the N proteins, the~M--N interaction induces the M3--M3 attractive interaction indirectly, which then provides the necessary curvature to trigger the budding process.  This can explain why for some coronaviruses, the~interaction of M--N is crucial for the formation of VLPs. Further experiments is necessary to understand if there is an attractive interaction between endodomains of M~proteins.  


\subsection{Condensation of RNA by N Proteins Is Essential for Its Packaging}

It is worth mentioning that the recent experiments on the SARS-CoV-2 M proteins show that the endodomains of the M proteins are highly positively charged, which might have an impact on their interactions with N and S proteins and RNA~\cite{Dolan2022}.  To~examine the effect of the M positive charges on the packaging of the genome, we also consider the case in which the endodomain (M3) has positive charges (+6, residues 101--222) (see Figure S7). Following~\cite{Collins2021}, we assume that the electrostatic repulsion between the M protein endodomains is screened out by the POPI lipids through the protein--lipid interactions, and~thus we set the M3--M3 electrostatic repulsion equal to zero.
Intuitively, one would expect that the positive charges on M proteins might be sufficient for the RNA packaging because of the attractive electrostatic interactions between RNA and the M endodomain. However, we find that the sole attractive electrostatic interaction between the M proteins and RNA is not enough for its packaging in the absence of N proteins, i.e.,~the condensation of genome by the N proteins is essential for the virus formation. This is in contrast to the case of many icosahedral viruses where the sole attractive interaction between the genome and the capsid proteins is sufficient for the encapsidation of the genome with viral capsid proteins~\cite{Li2017a,Siber-nonspecific}.  Because~of the length of the genome of coronaviruses and the budding process, the~presence of N proteins are essential for the genome condensation and virus formation. We have examined that, in our system, the~interaction between the positive charges on M and negative charges on RNA will be enough for budding if the length of the genome is much~shorter.

We note that all the budding simulations are performed using branched polymers with the number of branch points $N=30$.
Although the secondary structure of RNA does not have an impact on the final configurations of condensed RNPs, the~budding of a long linear chain is more challenging because the radius of gyration of a free linear chain is larger than a free branched polymer; thus, it takes much more time for the linear one to condense and be completely packaged. For~example for a linear chain with $L\sim800$, we have been able to package only half of the RNP during the time of our simulations (see Figure S8). 

\subsection{Impact of the Initial Distribution of M Proteins on Virus Budding}

\begin{figure}
    \centering
    \includegraphics[width=\linewidth]{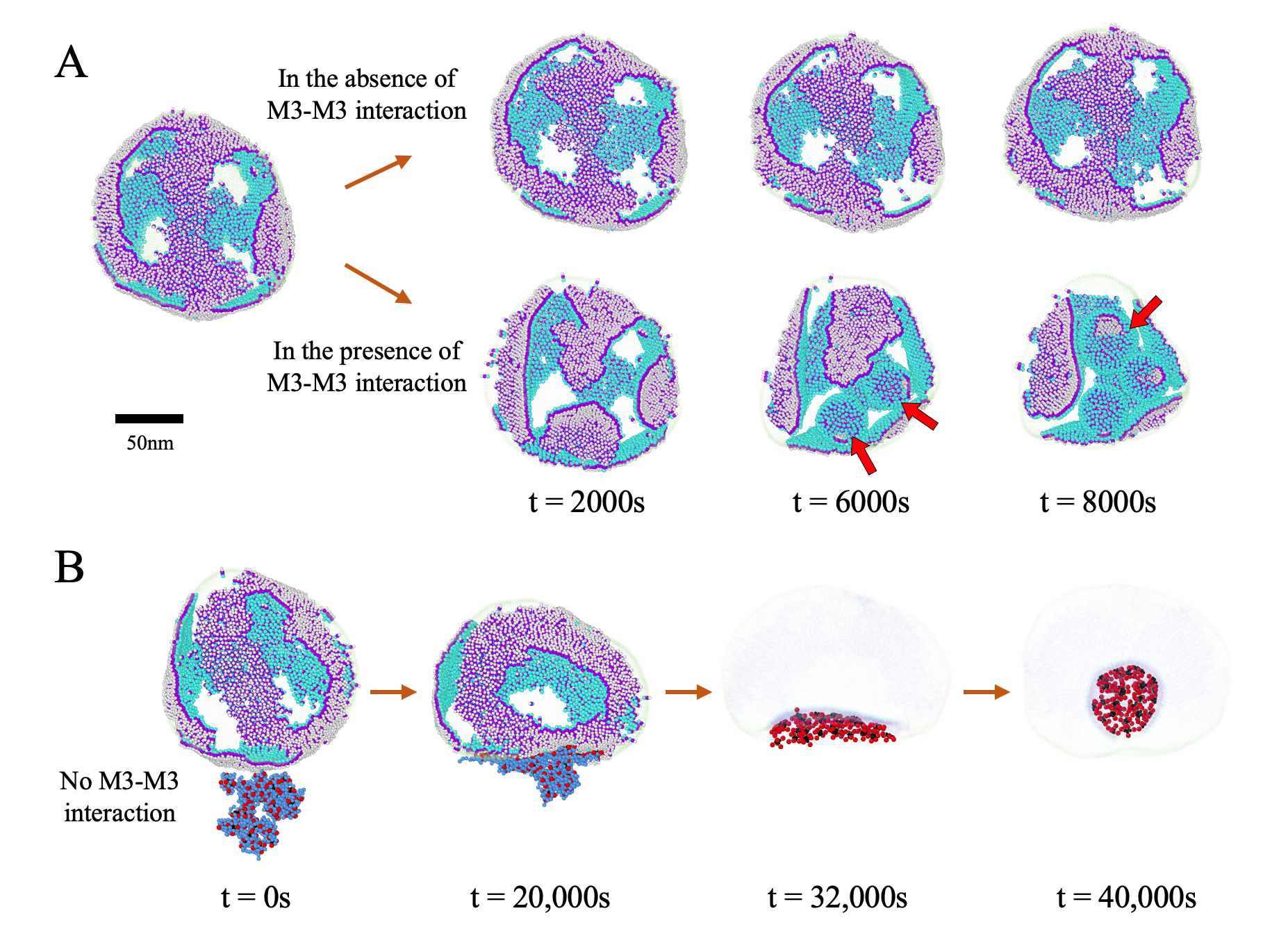}
    \caption{(\textbf{A}) {Snapshots} 
 of the simulations of the distribution of the M proteins embedded in the ERGIC membrane as a function of time. Initially, the~M proteins are randomly distributed in the ERGIC membrane. The~first snapshot at the left illustrates the distribution of the M proteins in ERGIC where the proteins and membrane have been relaxed for 16,000 s with $\epsilon_{M_2M_2}=1$ and $\epsilon_{M_3M_3}=0$. As~a result of diffusion, the~M proteins have formed a branched-like pattern. In~the absence of the M3--M3 interaction, the~branched-like pattern does not change (top row). However, with~increasing the endodomain--endodomain attractive interaction $\epsilon_{M_3M_3}$, the~connected patches of M proteins begin to aggregate into circular domains (bottom row, indicated with red arrows), which will eventually bud into empty vesicles. Note that the M proteins cover half of the surface of ERGIC, i.e.,~the M protein fraction is $f=50\%$ during the simulations; (\textbf{B}) snapshots of the simulations of the RNP budding from ERGIC. Initially, the~randomly distributed M proteins form a branched-like pattern on the surface of ERGIC in the absence of the endodomain interaction. However, the~interaction with RNP results in the aggregation of the proteins in the vicinity of RNP and the formation of the envelope enclosing RNP.  The~branched-like distribution is obtained by relaxing ERGIC and the M proteins for 16,000 s. The~RNP structure is obtained by relaxing N proteins (25 $\si{\micro M}$) and RNA for 8000 {s}. RNA is modeled as a branched polymers with the charge density $Z_g=-2$ and the number of branch points $N=30$. The~comparison of (\textbf{A},\textbf{B}) reveals that \hl{RNP} can mediate the interaction between the endodomains giving rise to a preferred curvature between M proteins. }
    \label{fig:random}
\end{figure}

We recall that, in the budding simulations presented above, we have always assumed that the M proteins initially form a spherical cap with a circular edge.  However, the~initial distribution of M proteins at the ERGIC membrane could be different.  To~this end, we have also considered the condition in which the initial distribution of the M proteins is random and have repeated the simulations presented in Figure~\ref{fig:ERGIC}A.d with the same parameters.  We first fully relax the ERGIC membrane and then allow the randomly distributed M proteins to diffuse and form different structures rather than imposing a circular cap from the beginning~Figure~\ref{fig:random}A.  As~we increase the strength of the interactions between the endodomain part of M proteins, the~irregular branched-like patterns start to form several circular domains pointed with arrows in Figure~\ref{fig:random}A, which will later bud into empty vesicles.  This disorder--order transition indicates that the spontaneous curvature of M proteins makes the branched configuration energetically unfavorable; as such, they diffuse to form separate circular~domains.  

Quite interestingly, we find that, even in the absence of the interaction between endodomains of M proteins, RNP can trigger the formation of circular caps (Video S8).  Figure~\ref{fig:random}B illustrates that, when we place RNP near the disordered regions where the M proteins embedded in ERGIC have formed a branched-like pattern, the~M--M interaction starts to bend the membrane due to the endodomain--RNP interaction. As~explained in the previous section, even in the absence of M3--M3 attractive interaction, the~interaction of RNP with the endodomain of M leads to an effective spontaneous curvature between M proteins.  The~curved region attracts more M proteins due to the energetically favorable endodomain--RNP interaction. After~a while, the~branched-like regions start to assume a circular configuration, which later rapidly buds through ERGIC encapsulating RNP to form the viral particle. We note that with a random distribution of M proteins, the~simultaneous packaging of genome and budding are not plausible through scenario 1 because then many N proteins basically follow the branched-like pattern of the M proteins and will be distributed everywhere on ERGIC. In~that case, RNA spreads out to reach those N proteins and thus it cannot condense to be packaged. Thus, a better understanding of the M--M interaction while embedded in the membrane can shed light on the pathway of genome condensation and its budding through the ERGIC~membrane.

\section{Discussion}
Even though the replication and assembly are conserved across many members of the family of coronaviruses, the~most rudimentary details of these processes remain elusive. This could be due to the presence of a number of structural proteins involved in the formation of viral particles and our limited knowledge regarding their specific roles.  Additionally, viral assembly and budding occur in the milieu of the intracellular host cell organelle membranes which further complicates isolating mechanisms specific only to the~virus.

In this paper, we have investigated the underlying physical principles for the formation of SARS-CoV-2 focusing on the role of the structural proteins in the assembly of the virus.  While many unknown parameters are involved in the packaging of the genome, our primary goal has been to obtain an overall understanding of under what conditions the large genomic RNA of coronaviruses can be condensed to bud through the host cell intercellular membrane. Specifically, we have sought to determine what types of effective protein--protein, genome--protein and membrane--protein interactions are necessary for the budding and formation of viral~particles. 

Despite all the experiments, the~quantitative values of the interactions critical to modeling remain experimentally unknown. To~this end, we have made several phase diagrams exploring how the relative strength of interactions between the different virus constituent parts influence the virus assembly and budding. Our simulations show that the essential requirement for the virus budding is the spontaneous curvature of the membrane proteins. Considering the length of SARS-CoV-2 RNA, under~no circumstances were we able to observe virus budding in the absence of the M protein spontaneous curvature.  We note that the attractive interaction between the cytosolic domain of M proteins or the presence of external scaffolding such as N proteins or RNP interacting with the cytosolic domain of M proteins can give rise to the spontaneous curvature of M proteins, see Figure~\ref{fig:random}.

The attractive interaction between the transmembrane domain of M proteins could also lower the budding energy barrier by introducing the line tension. A~very strong interaction between the transmembrane domains might create a high line tension, which could lead to virus budding. However, if~the M--M interaction is very strong, we observe that the M proteins form immobile large islands, trapped in metastable~configurations.

We have also investigated the mechanism of packaging of the long genomic RNA during the virus budding. Although~several experiments show that the N proteins and RNA phase separate~\cite{Cubuk2021,Savastano2020,Perdikari2020,Carlson2020,Wang2021,Wu2021}, the~exact location and when this process occurs remain unclear. 
 Our simulations offer two distinct plausible models: Either the RNP condensation and budding occur simultaneously at the ERGIC membrane or RNP is formed remotely and then transported to the ERGIC for budding. 
In both models, the~RNP forms a spherical layer under M proteins with a hollow center, consistent with previous observations~\cite{Klein2020}. In~the first scenario, however, the~order of arrival of N proteins and RNA to the ERGIC regions seems vital since the N proteins may initiate the budding process in the absence of genome and form empty particles (see Figure S9). If~the entire RNP is to be encapsulated, it will largely depend on the protein concentration, the~rate of RNP condensation, and the initial M protein~distribution.

In the alternative model that we offer in this paper, the~preformed RNP is transported to the ERGIC region and triggers the budding process by the M--N interaction. There will be no extra N proteins in the immediate vicinity of ERGIC in this model; thus, the~environmental factors have less impact on the budding process. 
However, if~the RNP configurations are specifically compact, in~the case of the strong N proteins--RNA interactions, the~packaging of the RNP will be difficult as the N proteins in the clusters need to reorient and expose their C-terminal domains to interact with M proteins. 
\revSL{Quite interestingly, a~recent paper on SARS-CoV-2 shows that N proteins aggregate around ER membranes that connect to viral RNA replication foci, whereas M, E, and~S proteins accumulate at the ERGIC membrane where the budding and maturation happen~\cite{Scherer2022}. Since N proteins are located near RNA, it is possible that RNP is pre-formed before meeting M, E, and~S proteins in ERGIC. Furthermore, the~study shows that N proteins and genome are observed at 5 h post infection (hpi), whereas M, E, and~S proteins are not observed until 7.5 hpi~\cite{Scherer2022}. The~fact that N proteins are expressed earlier than the other structural proteins indicates the possibility that RNA is condensed before being transported to the location of ERGIC. Considering these temporospatial aspects, we think that scenario 2 is completely plausible. However, more experiments are still needed to decipher where and when the genome is condensed.}
The comparison of the our simulations with the future in~vivo experiments focused on the kinetic pathways of RNP condensation can shed light on virus formation at the ERGIC~membrane.


\revSL{We note that, while some experiments regarding the assembly of both SARS-CoV and SARS-CoV-2 demonstrate that the minimal requirements for the virus budding are M and N proteins~\cite{Huang2004,Plescia2021}, other experiments show that M and E proteins are essential for the VLP formation~\cite{Mortola2004,Xu2020}. Interestingly, another set of experiments on SARS-CoV suggests that E proteins might not be required for the VLP assembly but rather enhance the production level~\cite{Siu2008,DeDiego2007}. Even with the unprecedented deluge of research the pandemic has wrought, the~role the structural proteins (E, M and N) play in the SARS-CoV-2 life cycle continue to remain speculative and pose an area of intense debate.}

\revSL{Our work can explain to some extent why there are inconsistencies in the experiments regarding the budding mechanism of various coronaviruses. Different requirements for budding could be due to the various strength of interactions between the structural proteins from one type of coronavirus to another, or~the composition of lipids in different host cells. In~one case, the~M--M interaction might be weak and the M--N interaction must be essential for the budding to occur, while in another case, the~presence of the E proteins confers a higher spontaneous curvature to the membrane~\cite{Ruch2012,Schoeman2019}; and as such, when E proteins are incorporated, a~significant enhancement of the viral particle production is observed~\cite{Siu2008}.}

Our aim is to shed light on the elusive assembly process of coronaviruses and particularly that of SARS-CoV-2, which could be the basis and guide for the design of future experiments too. A~systematic comparison of our simulations and the experiments with various mutations on M and N proteins could significantly advance our knowledge on this virus and the factors contributing to its formation. Understanding the complex life cycle of this virus inherently involves how the virus reproduces itself in such an extraordinarily successful way. Since there is always the danger of reduced effectiveness or even ineffectiveness of vaccines as a result of novel mutations, a~better understanding of the structural properties of SARS-CoV-2 could lead to alternative treatments for COVID-19 infections beyond the limited means we now~have.

\vspace{6pt}

\acknowledgments{R.Z. acknowledges support from NSF DMR-2131963, NSF DMR-2034794 and the University of California Multicampus Research Programs and Initiatives (Grant No. M21PR3267). S.L. acknowledges support from NSFC \hl{No.}12204335. We thank Paul van der Schoot for many helpful suggestions and Thomas Kuhlman for critically reading several sections of the paper.  We thank Yuanzhong Zhang, Ajay Gopinathan and Umar Mohideen for many helpful~discussions.}

\dataavailability{\hl{~~}All simulation data are present in the manuscript. HOOMD-blue plug-in is available at \url{https://github.com/virusassembly/hoomd_bondflip_plugin}.}

\reftitle{References}


\newpage

{
\fontsize{18}{18}\selectfont
\noindent\textbf{Supplementary Materials: Biophysical modeling of SARS-CoV-2 assembly: genome condensation and budding}
}

\setcounter{figure}{0}
\setcounter{section}{0}
\renewcommand{\thefigure}{S\arabic{figure}}

\section{Supplementary Figure}

\begin{figure}[H]
\includegraphics[width=\linewidth]{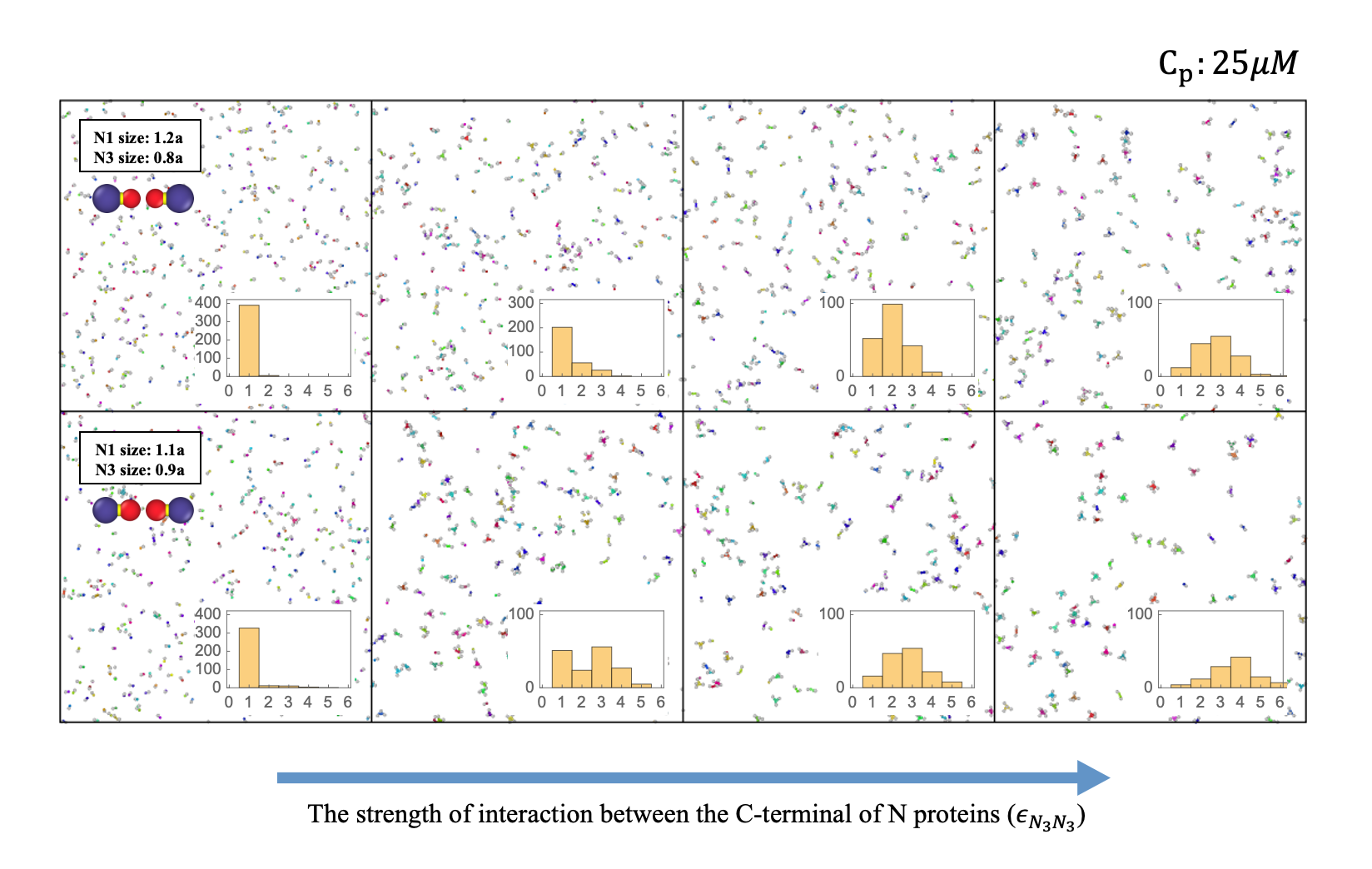}
\caption{
\textbf{Dimerization of the N proteins for various domain sizes and the strengths of interaction between the C-terminals.} The upper row shows N proteins with an N-terminal size of $1.2a$ and a C-terminal size of $0.8a$, where $a$ is the system length corresponding to $\SI{3}{nm}$. The size of N-terminal and C-terminal are $1.1a$ and $0.9a$, respectively, in the lower row. The size of the linker region N2 is kept constant and is $0.5a$. The N-terminals of N proteins are colored gray and the C-terminals of N proteins have different bright colors based on their cluster index (see SI Fig.~\ref{figsupp:RNPcluster} for the definition of the cluster index). The histograms in the figure show the count of the N protein oligomers. For the weak C-terminal interactions, we see that nearly all N proteins are monomers. As the interaction strength increases, more dimers and trimers form in solution. The interaction strengths $\epsilon_{N_3N_3}$ are $10$, $15$, $20$ and $50$ from left to right. We find that dimers are the dominant structure at $\epsilon_{N_3N_3}=20$.}
\label{figsupp:dimerphase}
\end{figure}

\newpage
\begin{figure}[H]
\includegraphics[width=\linewidth]{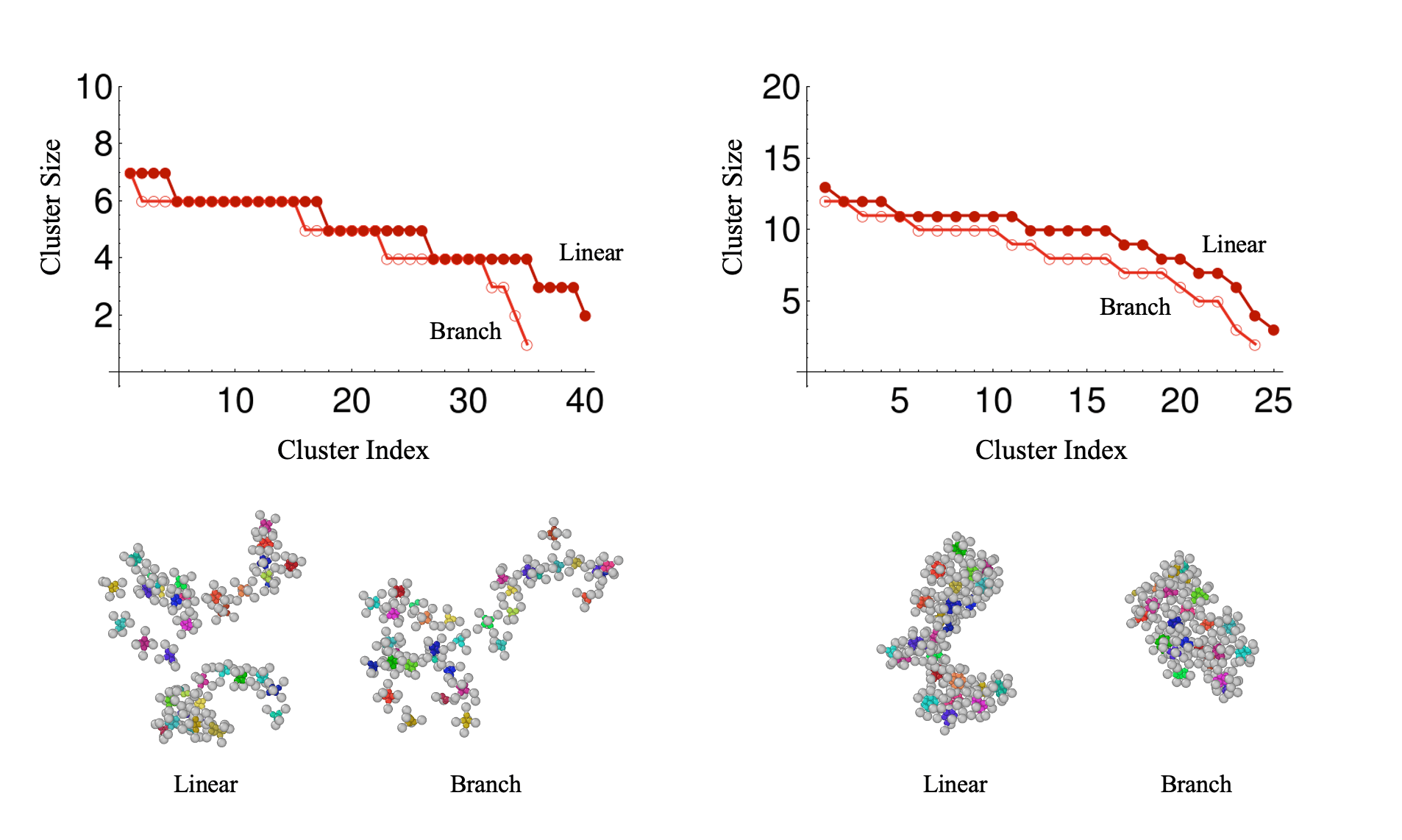}
\caption{\textbf{Cluster analysis of NPCs along RNP for weak (left) and strong (right) RNA-N protein interactions.} We recall that when the distance between the N-terminals of the N proteins are less than a cut-off distance, they are defined as a cluster assigned with a specific color and a unique cluster index. The plots display the number of N proteins in each cluster, where the clusters are sorted by their sizes. RNA is wrapped around the N proteins and is not shown in the figure.}
\label{figsupp:RNPcluster}
\end{figure}

\begin{figure}[H]
\includegraphics[width=\linewidth]{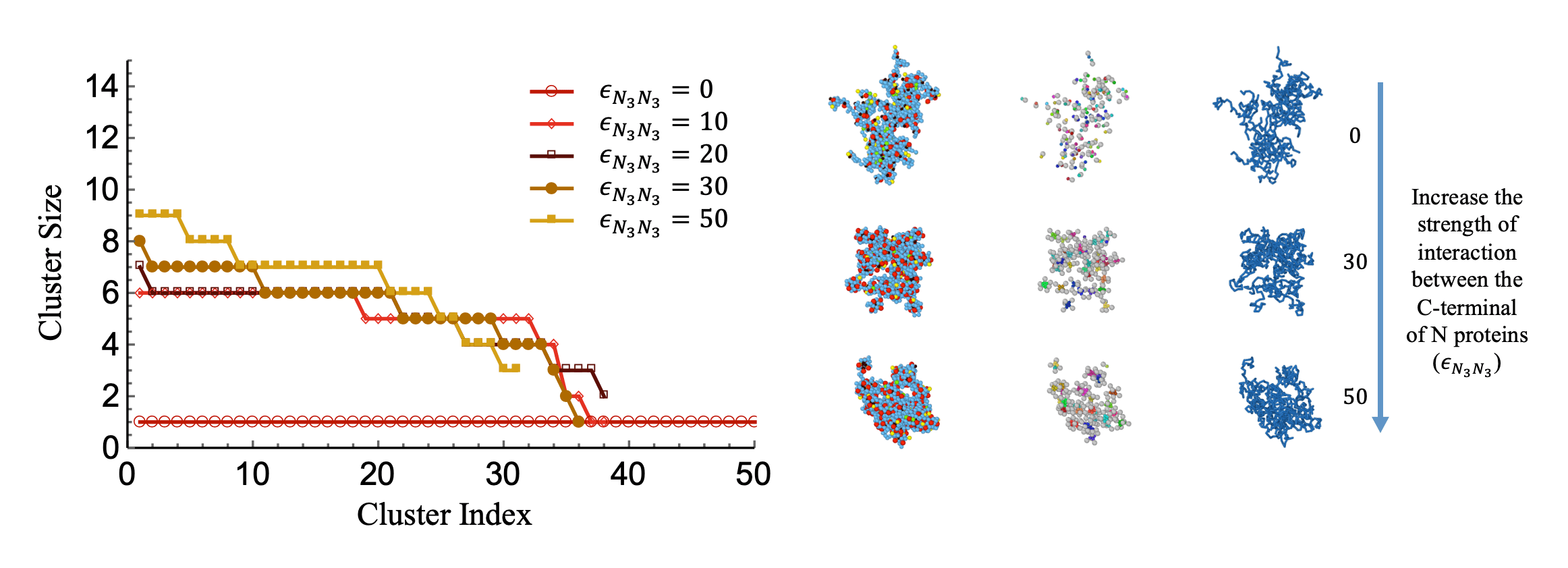}
\caption{\textbf{RNP condensation for various strengths of interactions between N proteins.} The left figure shows the cluster analysis for various N3-N3 interactions while the right figures display the structures of RNP, NPCs and RNA from left to right. The strength of interaction between the C-terminal of N proteins ($\epsilon_{N_3N_3}$) increases from top to bottom.  As the strength of the N-N interaction increases, the number of N proteins in each cluster increases, but the changes are small. Also, the size of RNP remains more or less the same. We note again that the N proteins in each cluster have the same color and a unique cluster index, and that the clusters are sorted by their sizes. The other parameters used are $Z_g=-2$, and N protein concentration is $25\si{\micro M}$.}
\label{figsupp:NNphase}
\end{figure}

\begin{figure}[H]
\includegraphics[width=\linewidth]{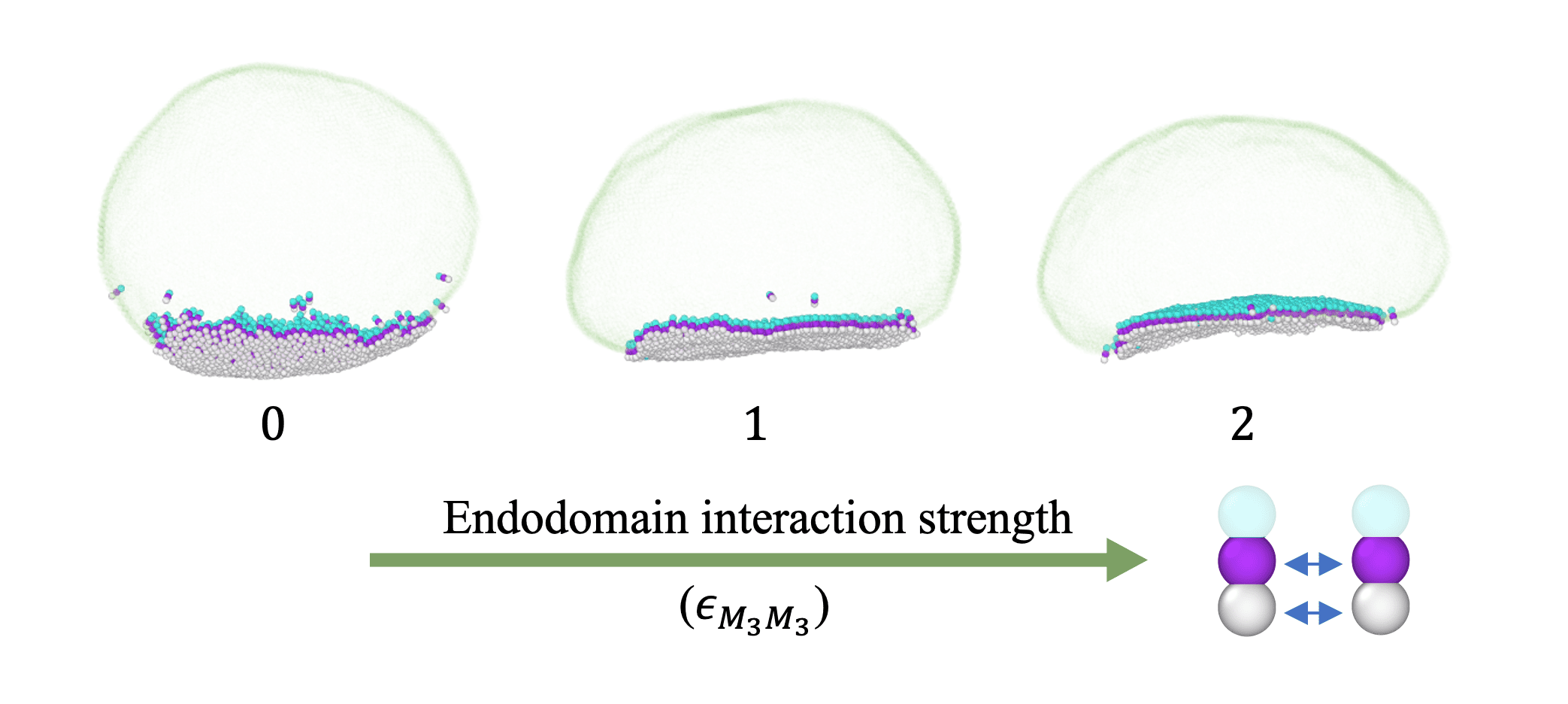}
\caption{\textbf{Illustration of the role of the spontaneous curvature in the budding process.}
The M protein has a cylindrical structure (no spontaneous curvature). All three spherical beads making the M protein have the same diameter. In particular, the transmembrane particle (purple) has the same size as endodomain particle (white). The transmembrane interaction strength is $\epsilon_{M_2M_2}=1$ and the endodomain (white) interaction strength $\epsilon_{M_3M_3}$ is $0$, $1$ and $2$ from top to bottom. There is no budding in the absence of spontaneous curvature.}
\label{figsupp:ERGICcyc}
\end{figure}

\begin{figure}[H]
\includegraphics[width=\linewidth]{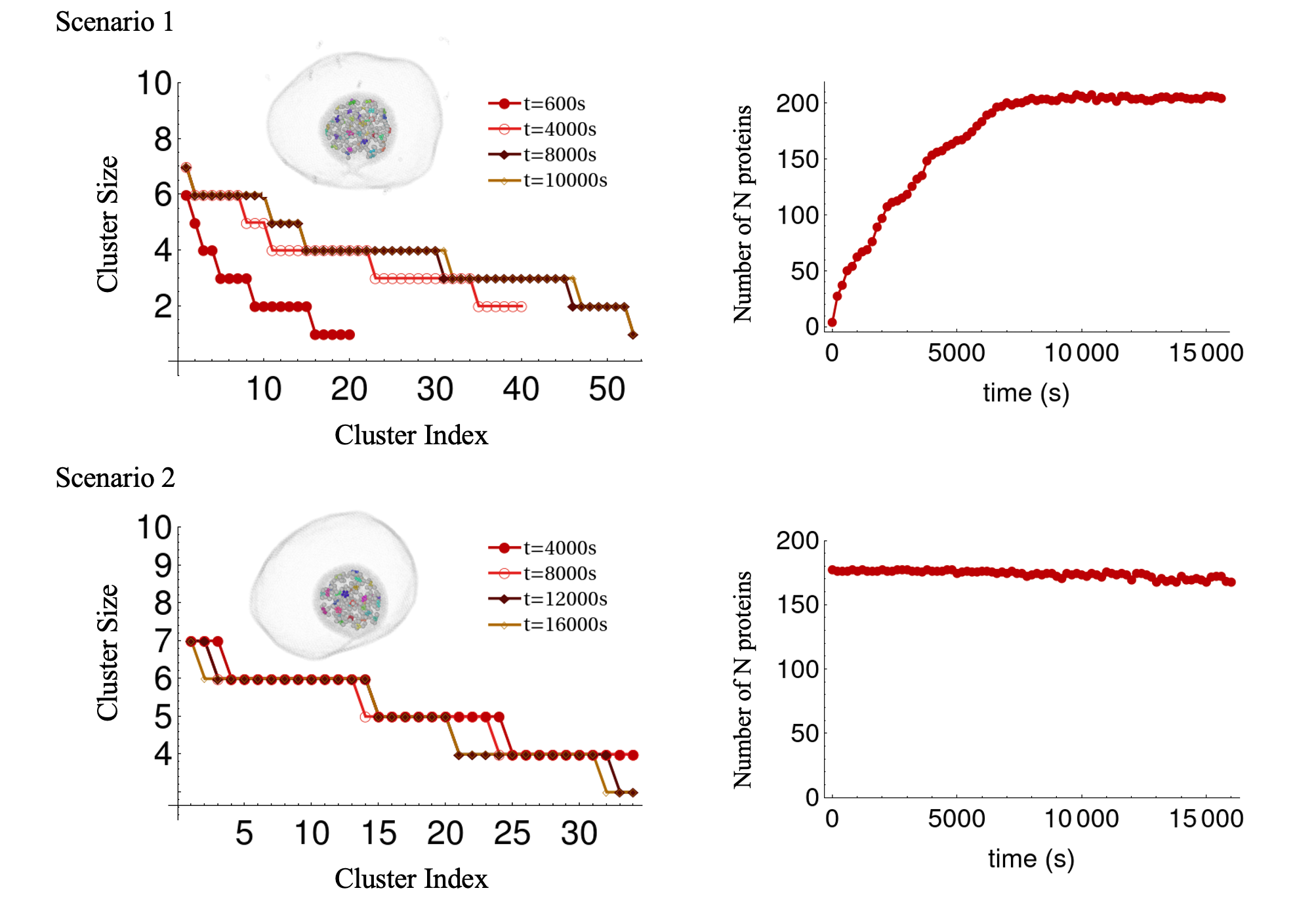}
\caption{\textbf{Cluster analysis of the N proteins along RNP during the budding.} The left figures show the number of N proteins in different clusters along RNP during the budding process vs.~the cluster index for scenario 1 (top) and scenario 2 (bottom). The insets show the configurations for which the clusters along RNPs were studied. Recall that the N proteins in the same cluster have the same color and index number, and that they are sorted by the size of the clusters.  The right figures show the total number of N proteins attached to RNA as a function of time during the budding process.  Since in the scenario 1, RNP condensation and budding take place at the same time, the number of N proteins increases as a function of time. However, in scenario 2, RNP has already formed and then has been transported near ERGIC before budding.}
\label{figsupp:cluster}
\end{figure}

\begin{figure}[H]
\includegraphics[width=\linewidth]{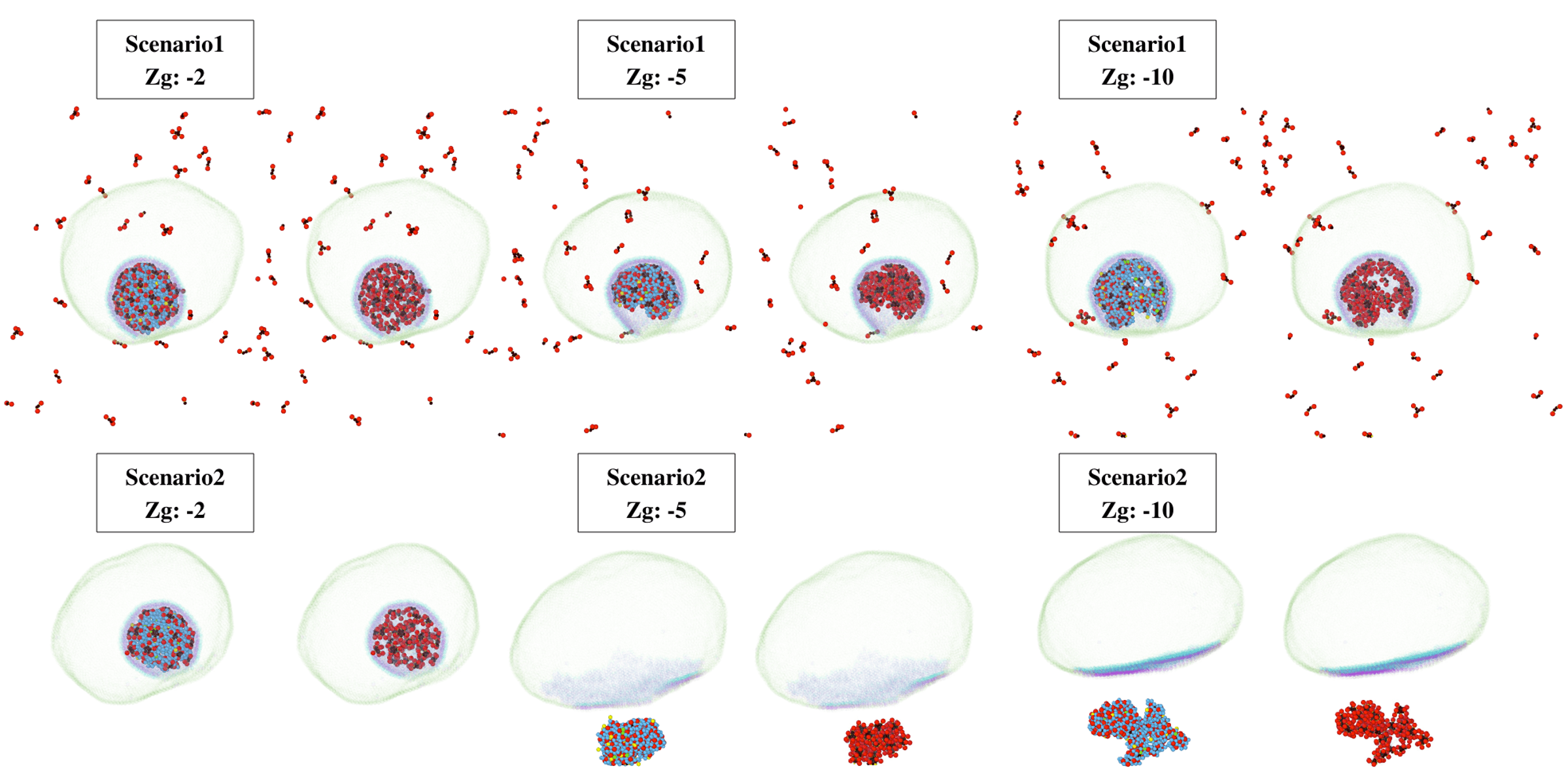}
\caption{
\textbf{RNP budding with various N protein-RNA interactions for scenarios 1 (top) and 2 (bottom) at $t=16,000s$.} There are three pairs of figures in each row. The left figure in each pair shows RNP and the right one shows the N proteins only. In scenario 1, ERGIC is initially located in the vicinity of RNA and the N proteins, and the RNA-N protein condensation and budding happen at the same time. In scenario 2, pre-condensed RNA-N proteins are placed near ERGIC, which later bud into a vesicle. See the assembly and budding movies SI Video S6 and Video S7. The figure also shows the simulations at various genome effective charge densities with $Z_g$ = $-2$, $-5$, $-10$ from left to right. We find that when the genome effective charge density is small, both scenarios (1 and 2) are plausible resulting in the formation of viral particles. However, as the genome effective charge density increases, more N proteins get absorbed to RNA, which makes it difficult for them to reorient and expose their C-terminal domains to interact with M proteins. }
\label{figsupp:phase}
\end{figure}

\begin{figure}[H]
\includegraphics[width=\linewidth]{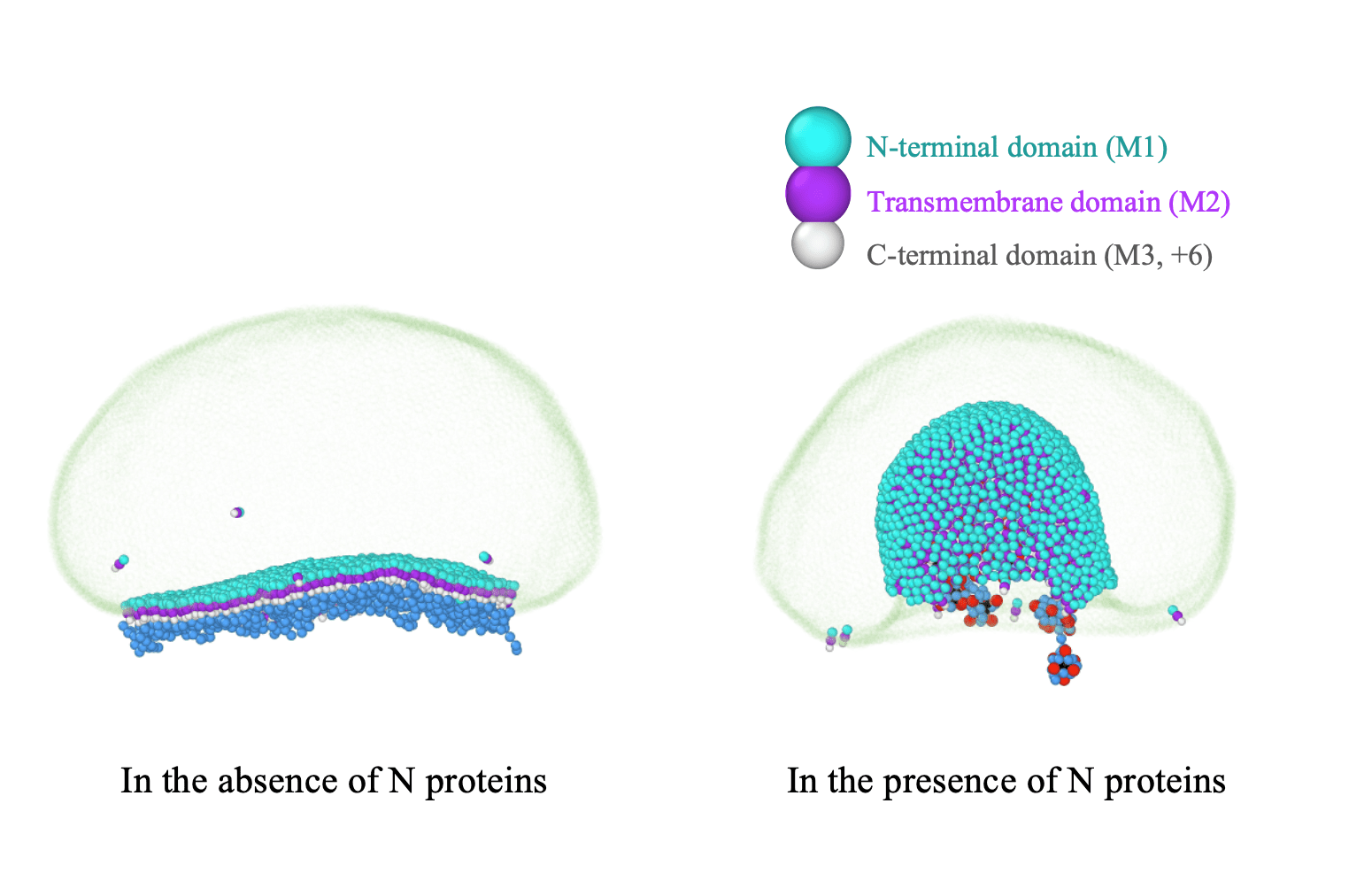}
\caption{\textbf{Snapshots of the simulations of the RNP budding when the endodomains of M proteins are positively charged.} The figure illustrates that the attractive electrostatic interactions between RNA and the M proteins is not sufficient for the RNA budding (left figure), and the presence of N proteins are essential for RNA packaging and budding (right figure). Note that a stronger N3-M3 interaction ($\epsilon_{N_3M_3}=20$) is required to trigger the budding compared to the neutral M protein case where a much weaker $\epsilon_{N_3M_3}=10$ results in the budding and assembly of the virus (Fig. 7). Both simulations are performed for $8000s$ and other parameters used are $L=800$, $\epsilon_{McMc}=1$, $\epsilon_{M_2M_2}=1$, $\epsilon_{M_3M_3}=0$, $Z_g=-2$ and $\epsilon_{N_3N_3}=20$.
We recall that the electrostatic repulsion between M3 and M3 is assumed to be screened and thus is zero. Note that for shorter chains, it's possible to package RNA in the absence of N proteins, but not relevant to the case of coronaviruses.}
\label{figsupp:Mcharge}
\end{figure}

\begin{figure}[H]
\includegraphics[width=\linewidth]{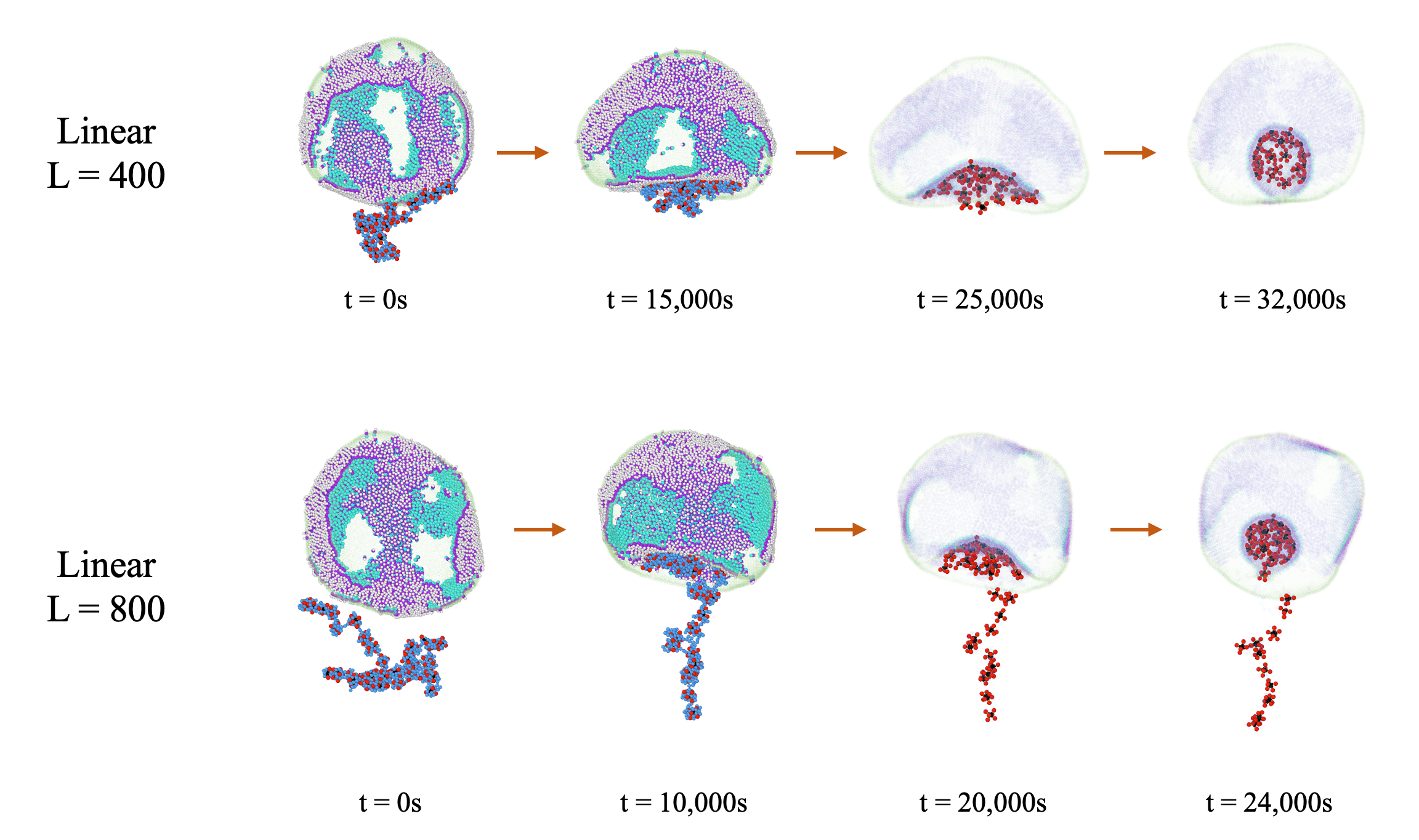}
\caption{\textbf{Snapshots of simulations of the RNP budding using a linear RNA.} Note, only shorter chains with $L=400$ (upper row) can be completely encapsidated. The M proteins are initially randomly distributed, and ERGIC has been fully relaxed. The $25\si{\micro M}$ protein concentration was used to condense RNA. The resulting RNP is then located near ERGIC. As illustrated in the figure, when $L=800$ (the bottom row), the chain cannot be fully packaged and a part of the chain will remain outside of the envelope. For the last two figures in each row, we show only the configuration of the N proteins. RNA has wrapped around the N proteins but is not shown in the figure. Note that RNP could be packaged if RNA were modeled as a branched polymer with the same length ($L=800$). The other parameters used are $\epsilon_{McMc}=1$, $\epsilon_{M_2M_2}=1$, $\epsilon_{M_3M_3}=0$, $Z_g=-2$, $\epsilon_{N_3N_3}=20$, and $\epsilon_{N_3M_3}=10$. }
\label{figsupp:linear}
\end{figure}

\begin{figure}[H]
\includegraphics[width=\linewidth]{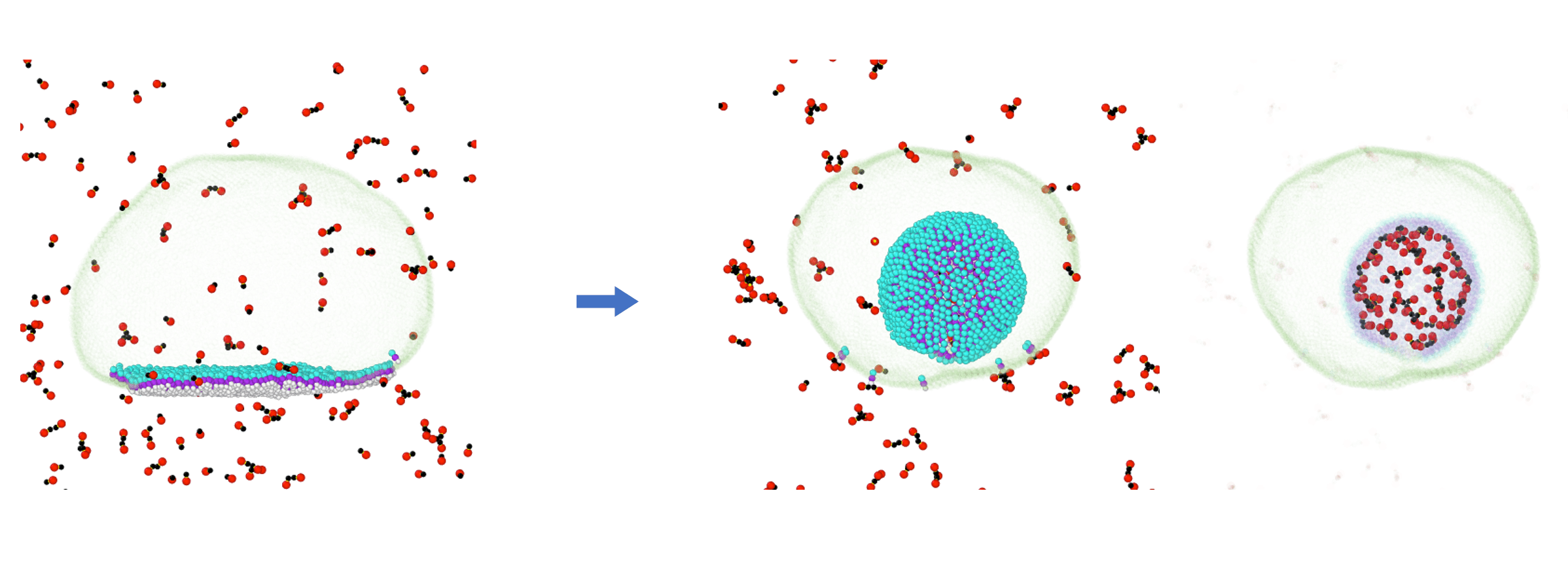}
\caption{\textbf{Formation of the virus like particles through the budding of N proteins in the absence of RNA.} The parameters used are $\epsilon_{M_2M_2}=1$, $\epsilon_{M_3M_3}=0$, $\epsilon_{N_3N_3}=20$, and $\epsilon_{N_3M_3}=10$. Note that in the absence of the N proteins, the M proteins cannot bud with the same M-M interaction strength parameters. }
\label{figsupp:VLP}
\end{figure}

\newpage
\setcounter{subsection}{0}
\renewcommand{\thesubsection}{Video S\arabic{subsection}}
\section{Supplementary Video}

\subsection{RNA condensation by the N proteins}\label{videosupp:condensation}

The N proteins of SARS-CoV-2 condense RNA. The concentration of proteins is $25\si{\micro M}$. RNA is modeled as a branched polymer with the number of branch points $N$ = $30$ and the genome effective charge density $Z_g$ = $-5$.

\subsection{The impact of the secondary structure of RNA on its condensation}\label{videosupp:branchN}

N proteins condense RNA with different number of branch points. The movies show RNP with the genome in blue and N proteins in red and black. The branch points are in green and the end points are in yellow. The total running time is $5000s$. The number of branch points from left to right is $N$ = $0$, $10$, $20$, and $30$.

\subsection{N proteins condense a linear genome for various N protein concentrations and effective genome charge densities}\label{videosupp:linear}

N proteins condense RNA modeled as a linear polymer for different protein concentrations and effective genome charge densities. The movies show RNP with genome in blue and N proteins in red and black. The total running time is $5000s$. The genome effective charge density increases from left to right with $Z_g$ = $-2$, $-5$, and $-10$.

\subsection{N proteins condense RNA at various N protein concentrations and effective genome charge densities}\label{videosupp:branch}

N proteins condense RNA modeled as a branched polymer at various protein concentrations and effective genome charge densities. The movies show RNP with RNA in blue and N proteins in red and black. The number of branch points is $N$ = $30$. The branched points are colored green and the end points are colored yellow. The total running time is $5000s$. The genome effective charge density increases from left to right with $Z_g$ = $-2$, $-5$, and $-10$.

\subsection{The role of the M protein interactions in budding at the ERGIC membrane}\label{videosupp:ERGIC}

Budding through the ERGIC membrane for various interaction strengths between the M protein transmembrane domains on the one hand and between the endodomains on the other hand. The lipid-lipid interaction strength is kept constant $\epsilon_{McMc} = 1$ for all simulations. The membrane stretching modulus is $k_s=20a^{-2}$ and the bending modulus is $k_b=20$. The total running time is $3000s$.

\subsection{RNP budding through the ERGIC membrane for scenario 1}\label{videosupp:budding1}

RNP budding through the ERGIC membrane for scenario 1. The N proteins interact with the RNA and form many clusters while simultaneously interacting with the M proteins embedded in the lipid membrane. Thus, the RNP condensation and budding occur simultaneously at the ERGIC membrane.

\subsection{RNP budding through the ERGIC membrane for scenario 2}\label{videosupp:budding2}
RNP budding through the ERGIC membrane for scenario 2. First, the N proteins interact with the RNA to form the condensed RNP complex, which later interacts with the M proteins that are embedded in the lipid membrane. Budding through the intercellular membrane then completes virus formation.

\subsection{RNP budding through the ERGIC membrane for the randomly distributed M proteins}\label{videosupp:buddingrandom}

RNP budding through the ERGIC membrane movie for the randomly distributed M proteins. See SI Video S7 for the budding through a circular cap made of M proteins.

\end{document}